\documentclass[12pt,a4paper]{article}

\usepackage{jheppub}
\usepackage{tensor}
\usepackage{amsmath}
\usepackage{amssymb}
\usepackage[colorlinks=true]{hyperref}
\usepackage{xcolor}
\usepackage{cancel}
\usepackage{dsfont}
\usepackage{extarrows}
\usepackage{todonotes}
\usepackage{physics}

\usepackage{titlesec}
\makeatletter
\g@addto@macro\bfseries{\boldmath}
\makeatother

\usepackage{framed}
\usepackage[most]{tcolorbox}
\usepackage{xcolor}
\colorlet{shadecolor}{gray!10}

\hypersetup{
    linkcolor=blue
}

\newcommand{\Z}{\mathbb{Z}}

\newcommand{\U}{\text{U}}

\newcommand{\Bock}{\text{Bock}}

\usepackage{tikz}

\usetikzlibrary{3d}
\usetikzlibrary{calc, decorations.markings}
\usetikzlibrary{decorations.pathmorphing}

\usepackage{diagbox}
\usetikzlibrary{positioning}
\usetikzlibrary{calc}
\usetikzlibrary{decorations.pathreplacing,calligraphy}

\usepackage{xstring}
\usetikzlibrary{decorations.pathmorphing} 
\usetikzlibrary{decorations.markings} 
\usetikzlibrary{arrows} 
\usetikzlibrary{shapes} 
\usetikzlibrary{matrix} 
\usetikzlibrary{positioning} 
\usepackage[english]{babel} 
\usepackage[autostyle]{csquotes}
\usepackage{pifont}
\usetikzlibrary{shapes.multipart}

\pgfkeys{/tikz/.cd, view angle/.initial=0, view angle/.store in=\picangle}

\tikzset{
    horizontal/.style={y={(0,sin(\picangle))}},
    vertical at/.style={x={([horizontal] #1:1)}, y={(0,cos(\picangle)cm)}},
    every label/.style={font=\tiny, inner sep=1pt},
    shorten/.style={shorten <=#1, shorten >=#1},
    shorten/.default=3pt,
    ->-/.style={decoration={markings, mark=at position #1 with {\arrow{>}}}, postaction={decorate}},
    ->-/.default=0.5,
    dark plane/.style={rotate around x=\angledp,canvas is xy plane at z=0,draw=cyan,fill=cyan!25},
}
\newcommand*{\Scale}[2][4]{\scalebox{#1}{$#2$}}

\title{Symmetries, anomalies, and dualities of two-dimensional Non-Linear Sigma Models}

\author{Guillermo Arias-Tamargo,}

\author{Maxwell L. Vel\'{a}squez Cotini Hutt}

\affiliation{Abdus Salam Centre for Theoretical Physics, The Blackett Laboratory, Imperial College London, Prince Consort Road, London, SW7 2AZ, UK}

\emailAdd{g.arias-tamargo@imperial.ac.uk}
\emailAdd{m.hutt22@imperial.ac.uk}

\abstract{We analyse the global symmetry structure of two-dimensional Non-Linear Sigma Models with Wess-Zumino term. When the target space has a compact isometry without fixed points, the theory has a pair of (group-like) global symmetries and many such theories also have non-invertible symmetries. We describe how the topology of the target space and Wess-Zumino term determine whether the group-like symmetries are continuous or discrete, and study their pure and mixed 't Hooft anomalies. We also revisit the construction of the non-invertible symmetries, which are associated with possible self-dualities under discrete gauging, and show how the global symmetry structure is left invariant by this gauging.
}

\begin{document}

\pagestyle{myplain}
\maketitle

\section{Introduction}

Non-invertible symmetries in two dimensions have been a major research focus of the past decade. Such symmetries correspond to the existence of a topological defect with no inverse --- that is, there is no defect with which it can be fused to yield the identity (see, e.g.~\cite{Bhardwaj:2023kri,Schafer-Nameki:2023jdn,Gomes:2023ahz,Sharpe:2015mja,Iqbal:2024pee,McGreevy:2022oyu,Cordova:2022ruw,Brennan:2023mmt,Luo:2023ive,Costa:2024wks,Shao:2023gho} for reviews). 
Early examples were found in \cite{Oshikawa:1996ww,Oshikawa:1996dj,Petkova:2000ip,Fuchs:2002cm,Fuchs:2003id,Fuchs:2004dz,Fuchs:2004xi,Frohlich:2004ef,Frohlich:2006ch,Fuchs:2007tx,Bachas:2012bj,Bachas:2013ora} and some of their applications were studied in \cite{Komargodski:2020mxz,Damia:2024kyt, Copetti:2024rqj,Copetti:2024dcz,Cordova:2024nux,Cordova:2024goh}. Meanwhile, formal developments include the identification of tensor categories as the suitable mathematical framework for their description \cite{Bhardwaj:2017xup,Chang:2018iay}.

A prototypical example of a non-invertible symmetry in two dimensions is the self-duality defect of the free compact boson \cite{Thorngren:2021yso,ChoiCordovaHsinLam2021,Niro:2022ctq,Damia2024,Bharadwaj:2024gpj,Argurio:2024ewp}. In this case, the field configuration corresponding to the defect can be constructed explicitly via the procedure of half-space gauging \cite{ChoiCordovaHsinLam2021}. 
This construction involves dividing the worldsheet into two regions and gauging a finite subgroup of a global symmetry in one of them. Thanks to T-duality, it is possible that the theories in both regions are equivalent, so the boundary between the two regions can be understood as a defect in a single theory. In the compact boson example there are two $\U(1)$ symmetries, momentum and winding, which we will denote by $\U(1)^m$ and $\U(1)^w$ respectively. If a $\Z_p$ subgroup of $\U(1)^m$ is gauged the theories in both regions are equivalent at a particular value of the radius: $2\pi R^2=p$. If instead a $\Z_p \times \Z_q$ subgroup of $\U(1)^m \times \U(1)^w$ is gauged (with $\gcd(p,q)=1$) a non-invertible defect can be found at any rational value of the radius.

As the theories in both regions of the worldsheet are required to be equivalent when the non-invertible defect is present, it must be the case that their global symmetries are the same. Demonstrating that this is the case, however, is non-trivial and relies on an interesting relation between group extensions and mixed anomalies discovered in \cite{Tachikawa:2017gyf}. 
The global symmetry of the compact boson is $\U(1)^m\times\U(1)^w$.\footnote{There is also a $\Z_2$ charge conjugation symmetry but we will not discuss it in the following.}
After gauging a $\Z_p$ subgroup of $\U(1)^m$ the theory has a dual (or quantum) $\Z_p$ global symmetry so, na\"{i}vely, it would seem that the global symmetry after gauging is
\begin{align}
    (\U(1)^m/\Z_p) \times \U(1)^w \times \Z_p\,,
\end{align}
which is not isomorphic to $\U(1)^m \times \U(1)^w$.
This reasoning neglects two important facts. First, the momentum and winding symmetries have a mixed 't Hooft anomaly; and second, $\U(1)^m$ is described by a non-trivial group extension by the $\Z_p$ subgroup being gauged (that is, $\U(1)^m/\Z_p \cong \U(1)$ but $\U(1)^m \ncong \U(1) \times \Z_p$).
The results of \cite{Tachikawa:2017gyf} can be used to infer the consequences of these two features under the gauging of the $\Z_p$ subgroup. Firstly, the mixed anomaly of momentum and winding results in a global symmetry in the gauged theory wherein the original $\U(1)^w$ symmetry is non-trivially extended by the dual $\Z_p$. That is, they do not form a product, $\U(1)^w \times \Z_p$, but rather combine into another $\U(1)$ symmetry which we denote $\U(1)^{\prime w}$.
Secondly, the fact that we are gauging a discrete subgroup involved in a non-trivial group extension causes the dual $\Z_p$ to have a mixed anomaly with the residual $\U(1)^m/\Z_p=\U(1)^{\prime m}$. All in all, both of these effects conspire so that the full symmetry structure after the gauging is again
\begin{align}
    \U(1)^{\prime m} \times \U(1)^{\prime w}\,,
\end{align}
and the two factors have a mixed anomaly.

The self-duality defect of the free compact boson has recently been generalised to a large class of Non-Linear Sigma Models (NLSMs) with Wess-Zumino (WZ) term in \cite{Arias-Tamargo:2025xdd} (see also \cite{Cordova:2023qei,Angius:2024evd,Caldararu:2025eoj} for discussions of other non-invertible symmetries in particular classes of two-dimensional NLSMs, and \cite{Chen:2022cyw,Chen:2023czk,Hsin:2022heo,Pace:2023kyi,Pace:2023mdo,Sheckler:2025rlk} for NLSMs in dimensions higher than two). These non-invertible defects are also constructed by gauging a discrete subgroup of a global symmetry in half-space and making use of T-duality. As for the compact boson, these symmetries exist provided that the NLSM satisfies a set of self-duality conditions. These generalise the constraint on the radius of the compact boson and require that the topology and geometry of the NLSM are preserved by the discrete half-space gauging.
The aim of the present work is to address the same question as above for this more general class of theories: how is the symmetry structure preserved by the discrete gauging? The answer again relies on the interplay between symmetry group extensions and mixed anomalies. 
For general NLSMs admitting a non-invertible defect, this information is encoded in two different short exact sequences of groups that are exchanged by the gauging. The same self-duality conditions that constrain the topology of the NLSM also ensure that these two sequences are the same, so that the symmetry structure is left unchanged by the gauging.

In order to tackle this problem, one first has to study in depth the global symmetries of NLSMs and their 't Hooft anomalies. In contrast to the compact boson, the symmetry structure of a generic NLSM with WZ term is not simply $\U(1) \times \U(1)$. We provide a detailed analysis of these issues. The global symmetry which generalises the momentum symmetry is associated with an isometry of the target space and, depending on the topology of the target space and the WZ term, may be either continuous or discrete. We will refer to it as the \emph{isometry symmetry}. In the continuous case, it has been studied in \cite{Hull:1989jk,HullSpence1991,Figueroa-OFarrill:1994uwr,Figueroa-OFarrill:1994vwl}, but the discrete case has not been discussed in as much detail and provides an interesting problem on its own.\footnote{The effect of topological terms in breaking the global symmetry to a discrete subgroup has been known in the context of quantum mechanics for some time \cite{Manton:1985jm} and analogous breaking in other dimensions has been discussed in \cite{Davighi:2018inx, Davighi:2020vcm}.} The global symmetry which generalises the winding symmetry, on the other hand, is more subtle. Winding symmetries are often associated with the fundamental group $\pi_1(M)$ (with $M$ being the target space of the NLSM). However, in the class of NLSMs that we consider here, this gives only a subgroup of the true global symmetry. The full structure is most conveniently described using T-duality and the symmetry generalising the winding symmetry of the compact boson is associated with an isometry of the target space of the T-dual NLSM. We will refer to it as the \emph{dual isometry symmetry}.

There is a beautiful geometric interpretation of the topology of the NLSM \cite{Hull2006} which can be used to directly infer the global symmetries of the model.
Namely, the topology is characterised by a pair of $\U(1)$ bundles and whether the isometry and dual isometry symmetries discussed above are continuous or discrete is determined by simple geometric properties of the curvature 2-forms associated with these bundles.
When the symmetries are continuous they may have pure 't Hooft anomalies, which also depend on the topology of the NLSM \cite{Hull2006}. In both cases there is a mixed 't Hooft anomaly between the two global symmetries.
We will compute the anomaly inflow action, which depends in a straightforward way on the curvature 2-forms.

The organization of this paper is as follows. We begin in section~\ref{sec:NLSMs} by reviewing the basics of the geometry of NLSMs, their isometry symmetry and its pure 't Hooft anomaly. In section~\ref{sec:T-duality}, we review the doubled torus construction and T-duality of two-dimensional NLSMs.
We then discuss the dual isometry symmetry, its anomaly, and the mixed anomaly with the isometry symmetry in section~\ref{sec:dual_isom}. Finally, in section \ref{sec:extensions} we put all the ingredients together and discuss how the discrete gauging preserves the symmetry structure when the self-duality conditions are satisfied. We conclude in section~\ref{sec:outlook} with a brief discussion. We include two appendices: appendix~\ref{app:quantisation} includes a technical derivation regarding the quantisation of the coefficient of the anomaly inflow action, and appendix~\ref{app:bockstein} collects some background material regarding discrete cochains and the Bockstein homomorphism required for the discussion of the mixed anomaly.

\section{Non-linear Sigma Models}
\label{sec:NLSMs}

We consider the following NLSM, which is a theory of maps $\Phi:W\to M$ with a WZ term,
\begin{align}\label{eq:NLSM_on_W}
    S = \frac{1}{2}\int_W g_{ij}\, dX^i\wedge \star dX^j + \frac{1}{2}\int_W b_{ij}\, dX^i\wedge dX^j\,,
\end{align}
where $X^i$ denote coordinates on the target space $M$, so the maps $\Phi$ are given locally by functions $X^i(\sigma)$ where $\sigma^a$ are the coordinates on the worldsheet $W$.
While $b$ locally defines a 2-form on $M$, globally it should be understood as a gerbe connection over $M$. The definition of the WZ term in \eqref{eq:NLSM_on_W} therefore needs clarification. 
In some cases we will find it useful to write this term instead as an integral on a three dimensional manifold $V$ such that $W=\partial V$,\footnote{The path integral does not depend on the choice of $V$ provided the flux is appropriately quantised such that $\frac{1}{2\pi}[H]$ defines an integral cohomology class.}
\begin{align}\label{eq:NLSM_action}
    S = \frac{1}{2}\int_W g_{ij}\, dX^i\wedge \star dX^j + \frac{1}{3!}\int_V H_{ijk} \, dX^i\wedge dX^j\wedge dX^k\,,
\end{align}
where $H=\dd b$. While all closed oriented 2-manifolds $W$ can be written as the boundary of an oriented 3-manifold $V$, the WZ term can only be written in this way if the map $\Phi:W\to M$ can be extended to a map $\widetilde{\Phi}:V\to M$ with $\widetilde{\Phi}\eval_{\partial V} = \Phi$. In particular, this is not possible if $\Phi(W)$ is a non-trivial 2-cycle in $H_2(M)$.
A definition of the WZ term purely in terms of $b$ is also possible using a generalisation of the Wu-Yang procedure \cite{Alvarez:1984es}.

The quantum theory does not depend on the choice of $V$ provided that $H/2\pi$ has integral fluxes on all 3-cycles in $M$ and so represents a class in $H^3(M,\Z)$. We will refer to the first term of \eqref{eq:NLSM_action} as $S_g$ and the second as $S_H$.

Suppose that $M$ has a compact isometry without fixed points, generated by a Killing vector $k$.\footnote{The case where $M$ has multiple commuting isometries can be treated similarly \cite{Hull2006}.} This implies that $M$ can be described as a circle bundle over some base $N$,
\begin{equation}\label{eq:M_bundle}
	S^1 \hookrightarrow M \twoheadrightarrow N \,.
\end{equation}
The topology of the bundle is determined by a $\U(1)$ connection $A$ on the base $N$, with curvature $F$. Let us denote the 1-form dual to $k$ by $\xi$, such that $\iota_k \xi = 1$. 
It has components $\xi_i = g_{ij} k^j / G$, where
\begin{equation}
    G = g_{ij} k^i k^j
\end{equation}
is the length-squared of the Killing vector. The metric on $M$ can be decomposed as
\begin{equation}\label{eq:metric_bundle}
    g = \bar{g} + G \xi \otimes \xi
\end{equation}
where $\bar{g}$ is a metric on the base $N$.

In what follows, we will choose local coordinates on $M$ denoted $X^i = (Y^\mu, X)$ where $Y^\mu$ are coordinates on the base $N$ and $X$ is a $2\pi$-periodic coordinate along the $S^1$ fibre, such that the Killing vector can be written locally as
\begin{equation}\label{eq:k_local}
	k = \partial_X \,.
\end{equation}
In these coordinates, we have
\begin{equation}\label{eq:xi=A+dX}
    \xi = A + \dd{X}
\end{equation}
and the curvature of the bundle \eqref{eq:M_bundle} is
\begin{equation}\label{eq:F_def}
    F = \dd A = \dd \xi \,.
\end{equation}
This satisfies $\iota_k F = 0$ and $\mathcal{L}_k F=0$, so it can be thought of as a form on the base $N$, rather than the total space $M$. A form with this property will be referred to as \emph{basic}.

\subsection{Isometry symmetry}
\label{subsec:isometry_symmetry}

Associated to the isometry of $M$, we can consider a transformation of the fields of the NLSM given by a shift along the Killing vector,
\begin{equation}\label{eq:symm_transf}
	\delta X^i = \alpha k^i
\end{equation}
for a constant $\alpha\in[0,2\pi)$. 
In the coordinates introduced above, this transformation is
\begin{equation}
	\delta Y^\mu = 0\qc \delta X = \alpha \,.
\end{equation}
This will be a global symmetry of the NLSM \eqref{eq:NLSM_action} provided that $g$ and $H$ are preserved by the Killing vector,
\begin{equation}\label{eq:preservation}
	\mathcal{L}_k g =0\,,\quad \mathcal{L}_k H = 0 \,.
\end{equation}
Under this transformation, the WZ term of the action \eqref{eq:NLSM_on_W} varies by
\begin{equation}\label{eq:WZ_variation}
	\Delta S = \alpha \int_W \Phi^*(\mathcal{L}_k b) = \alpha \int_W \Phi^*(\iota_k H) = \alpha \int_{\Phi(W)} \iota_k H \,,
\end{equation}
where we have used $\mathcal{L}_k b = \iota_k \dd{b} + \dd\iota_k b$ and that $W$ is closed. In the absence of the WZ term, therefore, the action is invariant and there is a $U(1)$ symmetry. With a non-vanishing WZ term, however, the symmetry is subtler.
It follows from \eqref{eq:preservation} and $\dd{H}=0$ that $\iota_k H$ is closed.
The nature of the global symmetry which results from the transformation \eqref{eq:symm_transf} depends on whether $\iota_k H$ is exact or not. We will refer to the symmetry as an \emph{isometry symmetry}. We will now show that when $\iota_k H$ is exact the isometry symmetry is $\U(1)$ and when $\iota_k H$ is not exact it is generically $\Z_\kappa$ for some $\kappa \in \Z_+$.

\subsubsection{U(1) symmetry: globally defined \texorpdfstring{$v$}{v}}
\label{subsec:U(1)isometry}

We first study the case where $\iota_k H$ is exact; that is, we can find a globally defined 1-form $v\in\Omega^1(M)$ such that
\begin{equation}\label{eq:iH_exact}
	\iota_k H = \frac{1}{2\pi} \dd v \,.
\end{equation}
Then, since $\partial W=0$, it follows that \eqref{eq:WZ_variation} vanishes by Stokes' theorem and the transformation \eqref{eq:symm_transf} is a symmetry of the action for any constant value of the parameter $\alpha$. That is, \eqref{eq:symm_transf} is a global $\U(1)$ isometry symmetry of the NLSM. It is important to remark that $v$ is then defined up to shifts by a closed 1-form; this will play a role often in the rest of the paper.

The 1-form Noether current associated with this symmetry is \cite{Hull2006}
\begin{equation}\label{eq:isom_current}
	j = g_{ij} k^j dX^i - \frac{1}{2\pi} v_i \star dX^i \,,
\end{equation}
which is conserved as a consequence of the equations of motion of \eqref{eq:NLSM_action}. Note that the current depends explicitly on $v$. From this, we can build a topological operator
\begin{equation}\label{eq:isom_top_op}
	U_\alpha (\gamma) = \exp(i\alpha \int_\gamma \star j)\,,
\end{equation}
where $\alpha \in [0,2\pi)$ and $\gamma$ is a 1-cycle in $W$. From Stokes' theorem and the conservation of $j$, $U_\alpha (\gamma)$ is unchanged by smooth deformations of $\gamma$ which do not cross charged insertions.

\paragraph{Example: \texorpdfstring{$M=S^3$}{M = S3}} Consider the NLSM with target space $S^3$ and non-zero H flux. We use Hopf coordinates and write the NLSM action with the same conventions as \cite{Arias-Tamargo:2025xdd} ($\phi,\psi\in[0,2\pi)\,,\,\theta\in[0,\pi]$),
\begin{align}
    S=\frac{R^2}{8}\int_{W} \left[d\theta^2 + \sin^2\theta \, d\psi^2 + 4\left( A+d\phi \right)^2 \right] + \frac{K}{4\pi}\int_{V} \sin\theta\, d\theta\wedge d\psi\wedge d\phi\,,
\end{align}
where $K \in\Z$ and
\begin{equation}\label{eq:S3_connec}
    A_{\pm} = \frac{1}{2} (\pm1-\cos\theta)\, \dd\psi 
\end{equation}
is the standard connection for the Hopf bundle. Here $\pm$ denote the two patches of the $S^2$ base. In particular, $+$ denotes the patch where the South pole is excluded, and $-$ denotes the patch where the North pole is excluded.
The transition function between the two patches is
\begin{align}\label{eq:North_South_poles_tf}
    \phi_+=\phi_- - \psi\,,
\end{align}
so the combination $\dd\phi + A$ is a globally defined 1-form on $M$.
In this example, the Killing vector is
\begin{align}
    k=\partial_\phi\,.
\end{align}
Looking at the WZ term, we can compute
\begin{align}
    \iota_k H = \frac{K}{4\pi}\, \sin\theta\, \dd\theta\wedge \dd\psi\,.
\end{align}
Since $H^2(S^3,\Z)=0$, this closed 2-form must be exact. Indeed, it can be written as in \eqref{eq:iH_exact} with
\begin{equation}\label{eq:v_global_S3}
    v = K ( A+\dd\phi )\,,
\end{equation}
which is globally defined.
We conclude that the NLSM with target space $S^3$ and $K$ units of H-flux has a $\U(1)$ global isometry symmetry. Note that the full global symmetry of the NLSM is generically larger than the $\U(1)$ isometry symmetry associated with the fibre. In this example, the $S^3$ target space has an SU(2) isometry group and the $\U(1)$ isometry symmetry we refer to is a particular $\U(1)$ subgroup associated with the $S^1$ fibre of the Hopf bundle. 

\subsubsection{Discrete symmetry: non-global \texorpdfstring{$v$}{v}}
\label{subsec:discrete_isometry}

Having understood the simple case where $\iota_k H$ is exact and there is a global $\U(1)$ isometry symmetry of the NLSM, let us turn to the subtler situation where $\iota_k H$ is not exact and ask whether there is a discrete remnant of the $\U(1)$ symmetry discussed above. Again, we look at the variation of the action \eqref{eq:WZ_variation}. Note that, since the action appears in an exponential in the path integral, it suffices to check that $\Delta S$ is equal to an integer times $2\pi$.

Since $H/2\pi$ defines an element of $H^3(M,\Z)$, the closed 2-form $\iota_k H$ also has integral periods and so defines an element of $H^2(M,\Z)$.\footnote{Recall that the Killing vector $k$ is normalised such that $k=\partial_X$ with $X$ a $2\pi$ periodic coordinate.}
As $\partial W=0$, the variation \eqref{eq:WZ_variation} is only sensitive to the cohomology class of $\iota_k H$.
Let us denote a basis of $H_2(M)$ by $\{\Sigma_i\}$, with $i=1,\dots,b_2(M)$ where $b_2(M)$ is the second Betti number of $M$. The integral $\int_{\Sigma_i} \iota_k H$ can be thought of as a map $H_2(M) \to\Z$ which assigns an integer to each 2-cycle $\Sigma_i$; that is,
\begin{equation}\label{eq:kappa_i_def}
	\int_{\Sigma_i} \iota_k H = \kappa_i \in \Z \,.
\end{equation}
The $\{\kappa_i\}$ are an equivalent way to describe the cohomology class $[\iota_k H] \in H^2(M,\Z)$.
Consider a field configuration $\Phi$ mapping $W \mapsto \Phi(W) = \Sigma_i \subset M$. The variation of the WZ term in this case evaluates to
\begin{equation}\label{eq:S_H_variation}
	\Delta S = \alpha \kappa_i
\end{equation}
and $\exp(iS_H)$ will be invariant provided that $\alpha \in (2\pi/\kappa_i)\Z$. Since $\alpha \in [0,2\pi)$, the transformation \eqref{eq:symm_transf} with these values of $\alpha$ corresponds to a discrete $\Z_{\kappa_i}$ shift symmetry. 
The path integral of the NLSM includes a sum over all maps $\Phi:W \to M$, and so includes contributions where $\Phi(W)$ wraps all 2-cycles in $M$. Therefore, the path integral will be invariant under discrete shifts where $\alpha \in (2\pi/\kappa)\Z$ with 
\begin{equation}\label{eq:kappa_gcd}
	\kappa = \abs{\text{gcd}(\kappa_i)} \,.
\end{equation}
Thus, the NLSM has a discrete $\Z_\kappa$ isometry symmetry (if $\kappa\neq0$). 
As noted below \eqref{eq:NLSM_action}, maps $\Phi$ for which $\Phi(W)$ is a homologically non-trivial 2-cycle in $M$ cannot be extended to maps $\widetilde{\Phi}:V\to M$. Hence this breaking of the $U(1)$ symmetry to a discrete subgroup by the WZ term can only be seen when these topologically non-trivial maps are included.

The values of the $\kappa_i$ can be found uniquely from the 3-form $H$, and so can be considered part of the topological data of the target space of the NLSM. In the case where $\iota_k H$ is exact, we have $\kappa_i=0$ for all $i$ and $\Delta S=0$ for all parameters $\alpha$, so we recover the $\U(1)$ symmetry of the previous subsection.\footnote{Generally, the isometry symmetry can be written as the Pontryagin dual group $\widehat{\Z}_\kappa \equiv \text{Hom}(\Z_\kappa,\U(1))$. This recovers the above results because for $\kappa\neq0$ we have $\widehat{\Z}_\kappa = \Z_\kappa$, and for $\kappa=0$ we have $\widehat{\Z}=\U(1)$.}

\paragraph{Example: \texorpdfstring{$M = S^2 \times S^1$}{M = S2 x S1}}

Let us describe a simple example of this type. Consider a product target space $M=S^2 \times S^1$. We denote the coordinate on the $S^1$ factor by $\phi$, so a Killing vector is $k=\partial_\phi$. The 3-form $H$ can be chosen as
\begin{equation}
	H = \frac{K}{4\pi} \sin\theta\, \dd\theta \wedge \dd\psi \wedge \dd \phi \,,
\end{equation}
where $\theta \in [0,\pi]$ and $\psi \in [0,2\pi)$ denote the standard coordinates on the $S^2$ factor. $H/2\pi$ has integral periods when $K\in\Z$. It is straightforward to check that \eqref{eq:preservation} is satisfied. In this case
\begin{equation}
	\iota_k H = \frac{K}{4\pi} \sin\theta\, \dd\theta \wedge \dd\psi \,,
\end{equation}
which is closed but not exact. This is clear since integrating it over the $S^2$ factor yields a non-zero result:
\begin{equation}\label{eq:int_iH_S2xS1}
	\int_{S^2} \iota_k H = K \,.
\end{equation}
A global 1-form $v$ satisfying \eqref{eq:iH_exact} does not exist in this case and there is no $\U(1)$ isometry symmetry. 
There is, however, a discrete symmetry which remains. In this case $b_2(M)=1$ and $H_2(M)$ is generated by the $S^2$ factor. Then, from \eqref{eq:S_H_variation} and \eqref{eq:int_iH_S2xS1}, the variation of the action under \eqref{eq:symm_transf} is $\Delta S=\alpha K$, so there is a surviving $\Z_{\abs{K}}$ isometry symmetry.

We remark that the key differences with the $S^3$ example discussed above are that 1) in the present example $M$ is a trivial $S^1$ bundle over $S^2$ so there is no non-trivial transition function mixing the fibre and base coordinates as in \eqref{eq:North_South_poles_tf}, and 2) the $S^2$ where we are integrating $\iota_k H$ is a \emph{bona fide} non-trivial 2-cycle in $S^2\times S^1$.\footnote{Generically, the base of a fibre bundle does not need to be a non-trivial cycle in the total space, or even a non-singular submanifold.}

\paragraph{Example: \texorpdfstring{$M=T^3$}{M = T3}}

Another simple example is to take a target space $M=T^3$, parametrised by coordinates $X,Y,Z\in[0,2\pi)$. The closed 3-form $H$ can be taken as
\begin{equation}
    H = \frac{K}{(2\pi)^2} \dd{X} \wedge \dd{Y} \wedge \dd{Z} \,,
\end{equation}
so $H/2\pi$ has integral periods when $K\in \Z$. Let us consider the isometry parametrised by the $X$ direction, i.e.~$k=\partial_X$. We then see the target space $T^3$ as a trivial $S^1$ bundle over $T^2$ with the base parametrised by the $Y,Z$ coordinates.
Then
\begin{equation}
    \iota_k H = \frac{K}{(2\pi)^2} \dd{Y}\wedge \dd{Z}
\end{equation}
is closed but not exact since
\begin{equation}
    \int_{T^2} \iota_k H = K \,.
\end{equation}
Therefore, there is no $\U(1)$ symmetry but there is a remnant $\Z_{\abs{K}}$ symmetry associated with the isometry.\\

In conclusion, in order to find the isometry symmetry of the NLSM associated to shifts along the fibre direction, one needs to determine whether $\iota_k H$ is exact or not. If it is, there is a U(1) symmetry. If it is not, there is a $\Z_\kappa$ symmetry, where $\kappa$ is specified by $H$ via \eqref{eq:kappa_gcd}.

\subsection{Topology of the NLSM}
\label{subsec:2bundles}

Let us consider in more detail the case where $\iota_k H$ is not exact. It follows from \eqref{eq:preservation} that $\iota_k H$ is closed, so it is locally exact. Denoting an open cover of $M$ by $\{M_a\}$, we can write
\begin{equation}\label{eq:iH_not_exact}
	\iota_k H = \frac{1}{2\pi} \dd v_a
\end{equation}
in a patch $M_a$. Note that this only defines the local 1-forms $v_a$ up to a shift by a closed 1-form. 
If it is possible to find $v_a$ which patch together to give a globally defined 1-form $v$ on $M$ then we return to the situation in subsection~\ref{subsec:U(1)isometry} where $\iota_k H$ is exact and there is a $\U(1)$ isometry symmetry. When $\iota_k H$ is not exact, the $v_a$ are not globally defined and the Noether current $j$ in \eqref{eq:isom_current} is not a well-defined operator \cite{Davighi:2018inx}.

Assuming that the $v_a$'s satisfy a cocycle condition, they can be understood as components of a $\U(1)$ connection $v$ over $M$, describing a larger target space
\begin{equation}\label{eq:Mhat_bundle}
    S^1 \hookrightarrow \hat{M} \twoheadrightarrow M \,.
\end{equation}
The curvature of the connection is\footnote{We will denote the connection by $v$, whose components in each patch $M_a$ are $v_a$. In the case where a globally defined $v$ exists, \eqref{eq:iH_not_exact} reduces to \eqref{eq:iH_exact}.}
\begin{equation}\label{eq:Ftilde_def}
    \widetilde{F} = \dd v =2\pi \iota_k H \,,
\end{equation}
which is basic. Recall from subsection~\ref{subsec:discrete_isometry} that $\iota_k H$ defines an element of $H^2(M,\Z)$. Then \eqref{eq:Ftilde_def} implies that the globally defined 2-form $\widetilde{F}$ has $2\pi\Z$ periods, as is required of the curvature of a conventionally normalised $\U(1)$ bundle. The 3-form $H$ can be decomposed into a basic 3-form $\bar{H}$ and a component involving the fibre direction as \cite{Hull2006}
\begin{equation}\label{eq:H_split}
    H = \bar{H} + \frac{1}{2\pi} \widetilde{F} \wedge \xi \,.
\end{equation}

From \eqref{eq:M_bundle}, $M$ itself is a $\U(1)$ bundle over $N$ and so $\hat{M}$ can be understood as a $T^2$ bundle over $N$ (assuming that the $\U(1)$ fibre of the bundle \eqref{eq:Mhat_bundle} fibres trivially over the $\U(1)$ fibre of the bundle \eqref{eq:M_bundle}),
\begin{align}\label{eq:Mhat_T2}
    T^2\hookrightarrow \hat{M}\twoheadrightarrow N\,.
\end{align}
The larger space $\hat{M}$ dictates the topology of the NLSM, and is specified by two $\U(1)$ bundles over $N$. One describes the target space $M$ as a $\U(1)$ bundle over a base $N$. Its curvature is $F$ in \eqref{eq:F_def}. The other is associated with the WZ term and describes a larger space $\hat{M}$ as a $\U(1)$ bundle over $M$. Its curvature is $\widetilde{F}$ in \eqref{eq:Ftilde_def}. 
The characteristic classes which characterise these bundles are $[F/2\pi]$ and $[\widetilde{F}/2\pi]$, which we will refer to as the \emph{Chern class} and \emph{H-class} respectively.

\subsection{'t Hooft anomaly of the isometry symmetry}

We now discuss the gauging of the isometry symmetry and its 't Hooft anomaly. We begin with the case where the isometry symmetry is $\U(1)$ and there is a globally defined 1-form $v$ before discussing the case where no such $v$ exists and there is a discrete $\Z_\kappa$ isometry symmetry.

\subsubsection{\texorpdfstring{$\U(1)$}{U(1)} symmetry: globally defined \texorpdfstring{$v$}{v}}
\label{subsec:anomalies_U(1)}

We begin with the case where $v$ is globally defined, so $\iota_k H$ is exact and we have a continuous $\U(1)$ isometry symmetry. The pure 't Hooft anomaly in this case has been discussed previously in \cite{Hull:1989jk,Jack:1989ne,Figueroa-OFarrill:1994uwr}.

A 't Hooft anomaly can be understood as an obstruction to gauging a global symmetry which can be compensated by coupling the theory to a topological theory in one dimension higher, referred to as the anomaly inflow theory. In our case, the anomaly inflow theory will be defined on $V$ such that its anomaly precisely cancels that of the NLSM on $W=\partial V$. The inflow action is known provided that $v$ can be chosen to satisfy
\begin{equation}\label{eq:Lv=0}
    \mathcal{L}_k v = 0 \,,
\end{equation}
which we will assume \cite{Hull:1989jk}.\footnote{One can always use the freedom to shift $v$ by a closed 1-form to ensure that this is satisfied locally \cite{Hull:2006va}, but it is not guaranteed to be possible globally.} In this case, it is known how to write the gauged theory on $V$ \cite{Hull2006}, and the action is given by
\begin{align}\label{eq:gauged_NLSM_action}
\begin{split}
    S_{\text{gauged}}&=\frac{1}{2}\int_W g_{ij}\, DX^i\wedge\star DX^j+\frac{1}{3!}\int_V H_{ijk} \, DX^i\wedge DX^j\wedge DX^k \\
    &\qquad +\frac{1}{2\pi}\int_V dC\wedge v_i DX^i\,,
\end{split}
\end{align}
where the covariant derivative is
\begin{align}
    D X^i=dX^i-C k^i\,.
\end{align}
The background field $C$ transforms simply as
\begin{equation}\label{eq:C_transf}
    C \to C + d\alpha \,.
\end{equation}
Note that the term in \eqref{eq:gauged_NLSM_action} involving the metric can simply be gauged by minimal coupling, whereas the WZ term requires the inclusion of an extra term.

In this situation, the 't Hooft anomaly can be interpreted as an obstruction to writing the action \eqref{eq:gauged_NLSM_action} on $W$. In other words, \eqref{eq:gauged_NLSM_action} already includes the inflow term. This can be seen explicitly as follows. We focus on the final two terms in \eqref{eq:gauged_NLSM_action} since the first is already written on $W$. Expanding the covariant derivatives and using that $\iota_k\iota_k H=0$, the final two terms of \eqref{eq:gauged_NLSM_action} can be written
\begin{align}
\begin{split}\label{eq:S_WZ_working}
    S_{\text{WZ, gauged}} &= \frac{1}{6}\int_V H_{ijk} \, dX^i\wedge dX^j\wedge dX^k - \frac{1}{2}\int_V(\iota_kH)_{jk}\, C\wedge dX^j\wedge dX^k\\
    &\qquad+\frac{1}{2\pi}\int_V dC\wedge v_i dX^i - \frac{1}{2\pi}\int_V dC\wedge (\iota_k v)C  \,. 
\end{split}
\end{align}
Using \eqref{eq:iH_exact} and that $H=\dd b$,
\begin{align}
\begin{split}
    S_{\text{WZ, gauged}} &= \int_V\Phi^*(\dd b)-\frac{1}{2\pi}\int_V C\wedge\Phi^*(\dd v)+\frac{1}{2\pi}\int_V dC\wedge \Phi^*(v) \\
    &\qquad - \frac{1}{2\pi}\int_V (\iota_k v) C\wedge dC\,.
\end{split}
\end{align}
Now, it follows from \eqref{eq:Lv=0} and $\iota_k\iota_k H=0$ that $\dd(\iota_k v)=0$, so $\iota_k v$ is a constant and
\begin{align}\label{eq:S_WZ_gauged}
    S_{\text{WZ, gauged}} = \int_W \Phi^*(b)+\frac{1}{2\pi}\int_W C\wedge \Phi^*(v)-\frac{\iota_k v}{2\pi}\int_V C\wedge dC\,.
\end{align}
Therefore, the action $S_{\text{gauged}}$ can be written
\begin{equation}\label{eq:Sgauged}
    S_{\text{gauged}} = S_{\text{gauged}}^W + S_{\text{inflow}} \,,
\end{equation}
where
\begin{equation}\label{eq:S_W_gauged}
    S^W_{\text{gauged}} = \frac{1}{2} \int_W g_{ij} DX^i \wedge \star DX^j + \int_W \Phi^*(b) + \frac{1}{2\pi} \int_W C \wedge \Phi^*(v)
\end{equation}
and the inflow action is
\begin{equation}\label{eq:pure_inflow}
    S_{\text{inflow}} = -\frac{\iota_k v}{2\pi} \int_V C \wedge dC \,,
\end{equation}
which is a three-dimensional Chern-Simons theory on $V$. In order for this to be well-defined on arbitrary 3-manifolds $V$, the coefficient $\iota_k v$ must be an integer. We demonstrate in Appendix~\ref{app:quantisation} that this is not an additional constraint on $v$, but is enforced by global consistency.

Neither of the terms on the RHS of \eqref{eq:Sgauged} are individually gauge-invariant, but the inflow term is designed to cancel the variation of the couplings to $C$ on $W$.
Therefore, unless $v$ can be chosen such that 
\begin{equation}\label{eq:iv=0}
    \iota_k v=0
\end{equation}
while also satisfying the previous constraints \eqref{eq:iH_exact} and \eqref{eq:Lv=0} then there is a 't Hooft anomaly of the $\U(1)$ isometry symmetry. In other words, there is no anomaly if $\iota_k H$ can be written as \eqref{eq:iH_exact} with $v$ a basic 1-form.
Equivalently, there is no 't Hooft anomaly if $\iota_k H$ is exact not only as a 2-form on $M$ but also as a 2-form on the base $N$.

As always when discussing 't Hooft anomalies, we must check that there is no choice of local counter-term which can be added to the theory on $W$ which removes the anomaly. If such a counter-term exists, then a two-dimensional coupling to the background field $C$ can be found such that the path integral is invariant under background gauge transformations and there is no anomaly. For continuous symmetries, the ability to add counter-terms is related to the fact that the Noether current is not uniquely defined. In particular, we are free to add any co-closed 1-form to a Noether current without spoiling its conservation. In the present case, the Noether current associated with the isometry symmetry is given in \eqref{eq:isom_current}. When gauging the symmetry, the linear coupling to the background gauge field is proportional to $C \wedge \star j$ and the addition of a co-closed 1-form to the Noether current simply corresponds to a different choice of coupling to the background field $C$ (i.e.~the addition of a local counter-term). 

The Noether current \eqref{eq:isom_current} depends explicitly on the 1-form $v$. This 1-form is not unique, and can be shifted by a closed 1-form $w$ with $\mathcal{L}_k w=0$ without spoiling the properties \eqref{eq:iH_exact} and \eqref{eq:Lv=0}. Shifting $v$ in this manner has the effect of changing the Noether current:
\begin{equation}
    j \to j - \frac{1}{2\pi} \star \Phi^*(w) \,.
\end{equation}
Since $w$ is closed, this extra term is co-closed and this modification of the Noether current does not affect its conservation. One can see explicitly from \eqref{eq:S_WZ_gauged} that shifting $v$ in this way introduces a new counter-term linear in $C$ in the gauged theory and changes the coefficient of the inflow theory.

The anomaly will vanish if we can shift $v$ such that \eqref{eq:iv=0} is satisfied. We now discuss whether this is possible. In the coordinates $X^i = (Y^\mu, X)$, we have
\begin{equation}\label{eq:v=v'+fibre}
    v = v' + (\iota_k v) \dd{X} \,,
\end{equation}
where $v'$ is basic. 
Let us first consider the simple case where $M=N\times S^1$, i.e.~when the bundle \eqref{eq:M_bundle} is trivial. In this case the bundle admits a global section and $w=-(\iota_k v)\dd{X}$ is a globally defined closed 1-form on $M$ with $\mathcal{L}_k w = 0$. Therefore, we can shift $v$ in \eqref{eq:v=v'+fibre} by $w$ to give $v'$, which has $\iota_k v'=0$. Since both $v$ and $w$ are globally defined, so is $v'$. Using $v'$ instead of $v$, the inflow term \eqref{eq:pure_inflow} vanishes and the symmetry has no 't Hooft anomaly. 

As stressed above, this can be seen simply as a redefinition of the Noether current or, equivalently, the introduction of a local counter-term. Let us explore how this works in this example. The inflow action \eqref{eq:pure_inflow} expresses the fact that under a transformation \eqref{eq:symm_transf} the partition function $Z[C]$ of the two-dimensional gauged theory (without coupling to the inflow theory) varies as
\begin{equation}\label{eq:Z[C]_variation}
    Z[C+d\alpha] = \exp(\frac{i}{2\pi} (\iota_k v)\int_W C \wedge d\alpha) Z[C]\,.
\end{equation}
Changing $v$ to $v'$ as above has the effect of adding to the gauged theory \eqref{eq:S_WZ_gauged} a new linear term
\begin{equation}
    - \frac{\iota_k v}{2\pi} \int_W C \wedge dX 
\end{equation}
and removing the inflow term.
Under the action of the symmetry, the variation of this new linear term is
\begin{equation}
    -\frac{\iota_k v}{2\pi} \int_W \left(C \wedge d\alpha + d\alpha \wedge dX \right) = -\frac{\iota_k v}{2\pi} \int_W C\wedge d\alpha + 2 \pi \Z
\end{equation}
where we used that $d\alpha \wedge d\alpha=0$, that both $\alpha$ and $X$ are $2\pi$-periodic, and also that $\iota_k v \in\Z$ (see Appendix~\ref{app:quantisation}). Therefore, in the path integral, the variation of this term precisely cancels the variation \eqref{eq:Z[C]_variation} and the resulting theory is gauge-invariant. As the inflow term vanishes, the gauged theory is purely two-dimensional. We conclude that when $M=N\times S^1$ we can always shift $v$ in such a way that $\iota_k v=0$ and there is no 't Hooft anomaly.

Suppose now that the bundle \eqref{eq:M_bundle} is non-trivial, so $M$ is not just a product space. Na\"{i}vely, the operations above would still hold, but there is an important subtlety. Since we are considering the case where $\iota_k H$ is exact, we must choose a globally defined $v$ in order to give a $\U(1)$ isometry symmetry. If $M$ is a non-trivial bundle then it does not admit a global section and $w=-(\iota_k v)\dd{X}$ is no longer a globally defined closed 1-form. This can be seen clearly as follows.
Since $M$ is a non-trivial $S^1$ bundle over the base $N$, on overlaps of open patches in the base the fibres are connected by transition functions: $X \to X - f$ (for some $f\sim f+2\pi$). Since $v$ in \eqref{eq:v=v'+fibre} is globally defined, $v'$ must transform as a connection $v' \to v' + (\iota_k v) \dd{f}$. That is, neither $v'$ nor $\dd{X}$ is a globally defined 1-form, but the combination \eqref{eq:v=v'+fibre} is. 

Given this, we cannot simply shift $v$ by $w$ because the result, $v'$, is not a globally defined 1-form.
Instead, we could try to shift $v$ by $w=w'-(\iota_k v)\dd{X}$ where $w'$ is chosen such that $w$ is globally defined. As above, this requires $w'$ to transform as a non-trivial $\U(1)$ connection (i.e.~with non-zero curvature). However, we are only allowed to shift $v$ by a closed 1-form, so we must demand $\dd{w} = \dd{w'} = 0$ which is a contradiction. Therefore, there is no globally defined closed 1-form which we can shift $v$ by which changes the value of $\iota_k v$. It follows that, if $M$ is a non-trivial bundle, the $\U(1)$ isometry symmetry has a 't Hooft anomaly \eqref{eq:pure_inflow} and the prefactor $\iota_k v$ is independent of the choice of globally defined 1-form $v$.

A simple example of this type is $M=S^3$, which was discussed in section~\ref{subsec:U(1)isometry}. There, the fibre coordinate was denoted $\phi$ instead of $X$ and there are non-trivial transition functions \eqref{eq:North_South_poles_tf}. A globally defined choice of $v$ was given in \eqref{eq:v_global_S3}. Shifting $v$ by $-K\dd\phi$ sets $\iota_k v=0$ but the resulting $v'$ is not globally defined.

So far we have found that when $\iota_k H$ is exact the $\U(1)$ isometry symmetry does not have a 't Hooft anomaly if $M$ is a trivial bundle, but has an anomaly \eqref{eq:pure_inflow} if $M$ is a non-trivial $S^1$ bundle and $\iota_k v\neq0$.
In the latter case, while the full $\U(1)$ isometry symmetry has a 't Hooft anomaly, there may be a discrete subgroup which is non-anomalous.
The inflow \eqref{eq:pure_inflow} implies that under a transformation \eqref{eq:symm_transf} the partition function $Z[C]$ of the gauged theory varies as
\begin{equation}\label{eq:Z[C]_variation2}
    Z[C+d\alpha] = \exp(\frac{i}{2\pi} (\iota_k v)\int_W \alpha\, dC) Z[C]\,.
\end{equation}
This is equivalent to \eqref{eq:Z[C]_variation} by integrating by parts.
Thus, when the parameter $\alpha$ takes constant values $\alpha \in (2\pi/\iota_k v)\Z$, the partition function is invariant. These values of the parameter $\alpha$ generate a non-anomalous $\Z_p$ subgroup of the $\U(1)$ isometry symmetry with 
\begin{equation}\label{eq:p_def}
    p = \abs{\iota_k v} \,.
\end{equation}

\paragraph{Example: \texorpdfstring{$M=S^3$}{M = S3}}

Let us briefly discuss the example of $M=S^3$ discussed in subsection~\ref{subsec:U(1)isometry}. In this example $\iota_k H$ is exact so there is a $\U(1)$ isometry symmetry and a globally defined choice of $v$ was given in \eqref{eq:v_global_S3}. This $\U(1)$ symmetry has a 't Hooft anomaly since $p=\abs{\iota_k v}=\abs{K} \neq0$. There is a non-anomalous $\Z_{\abs{K}}$ subgroup.

\subsubsection{Discrete symmetry: non-global \texorpdfstring{$v$}{v}}
\label{subsec:anomalies_discrete}

In the case where $\iota_k H$ is not exact and there is a discrete $\Z_\kappa$ isometry symmetry, it is less clear how to couple background fields. In some applications, given a discrete $\Z_n$ symmetry, one can look to embed it in a larger $\U(1)$ symmetry, compute the $\U(1)$ anomaly using standard methods and then restrict the background field to represent only a $\Z_n$ subgroup. In the present case, it is not clear how to implement this approach because the action \eqref{eq:gauged_NLSM_action}, where the $\U(1)$ symmetry is gauged, explicitly depends on $v$. When $\iota_k H$ is not exact, there is no globally defined $v$ satisfying \eqref{eq:iH_exact} so this action is not well-defined (i.e.~it is not invariant under transformations $v\to v+\dd{f}$). We will return to an adaptation of the method used above for the $\U(1)$ symmetry in the following section. The idea will be to introduce a continuous $\U(1)$ background gauge field as well as a Lagrange multiplier. The effect of the latter is to restrict the gauge field to a discrete $\Z_n$ subgroup of $\U(1)$, so only the finite symmetry is gauged. We will see that allowing the Lagrange multiplier itself to transform under the symmetry can result in a gauge-invariant coupling to the background field for the discrete symmetry. The global structure of the couplings of the background field and Lagrange multiplier can be understood geometrically as an extension of the target space.

\section{Doubled torus construction}
\label{sec:T-duality}

In this section, we review the derivation of \cite{Hull2006} of T-duality of NLSMs whose target space has a compact isometry without fixed points. This is done by finding a gauged NLSM which is equivalent to the original one after gauge fixing. The gauged NLSM includes some auxiliary fields such that, by swapping the order of path integration of the fields, the path integral of the gauged NLSM is also equivalent to that of a NLSM with a different (T-dual) target space. As in previous sections, we will consider only a single isometry but the extension to multiple Abelian isometries is simple \cite{Hull2006}.

Geometrically, the gauged NLSM is constructed by enlarging the target space $M$ of the original NLSM to the space $\hat{M}$ in \eqref{eq:Mhat_T2}. As discussed in subsection~\ref{subsec:2bundles}, $\hat{M}$ is a $T^2$ bundle over a base $N$. We denote the coordinates on $\hat{M}$ by $\hat{X}^I=\{Y^\mu,X,\hat{X}\}$, where $Y^\mu$ are coordinates on the base $N$, $X$ is a coordinate on the first $\U(1)$ fibre and $\hat{X}$ is a coordinate on the second $\U(1)$ fibre. Both $X$ and $\hat{X}$ are taken to be $2\pi$-periodic. The fibration of the $X$ direction over the base is described by the curvature $F$ and the fibration of the $\hat{X}$ direction is described by $\widetilde{F}$. Since the dimension of the fibre of this bundle is twice as large as the original bundle \eqref{eq:M_bundle}, this is referred to as the \emph{doubled torus construction}. We begin by discussing the isometry symmetry of the NLSM with target space $\hat{M}$ and then construct the gauged NLSM which is equivalent to the original NLSM on $M$ in the next subsection.

The metric $\hat{g}$ and 3-form $\hat{H}$ on $\hat{M}$ are taken to be
\begin{equation}\label{eq:ghat}
    \hat{g}_{ij} = g_{ij} \qc \hat{H}_{ijk} = H_{ijk} \,,
\end{equation}
with all other components vanishing. Since $g$ and $H$ satisfy \eqref{eq:preservation}, we have $\mathcal{L}_k \hat{g} = \mathcal{L}_k \hat{H}=0$ also. Furthermore, the transition functions of $\hat{X}$ can always be chosen such that
\begin{align}\label{eq:vhat}
    \hat{v}=v+\,\dd\hat{X}\,,
\end{align}
is a globally defined 1-form. Clearly, $\dd\hat{v}=\dd{v}$, so we have
\begin{equation}
    \iota_k \hat{H} = \frac{1}{2\pi} \dd\hat{v} \,.
\end{equation}
That is, $\iota_k \hat{H}$ is \emph{always} exact on $\hat{M}$. Therefore, the NLSM with target space $\hat{M}$ always has a $\U(1)$ isometry symmetry associated with the Killing vector $k$:
\begin{equation}
    \delta \hat{X}^I = \alpha k^I \,.
\end{equation}
We emphasise that this is the case even if $\iota_k H$ is not exact on $M$, so the original NLSM on $M$ only has a discrete isometry symmetry. The path integral of the gauged NLSM (which is equivalent to that of the original NLSM) is found by gauging this symmetry, as we discuss in the next subsection.

We also note that the NLSM on $\hat{M}$ has another isometry given by
\begin{equation}\label{eq:ktilde}
    \widetilde{k} = \partial_{\hat{X}} \,.
\end{equation}
Since $\hat{g}$ has no dependence on the $\hat{X}$ coordinate, this is a Killing vector. The 1-form dual to $\widetilde{k}$ is 
\begin{equation}
    \widetilde{\xi} = \hat{v}
\end{equation}
since, from \eqref{eq:vhat}, we have $\iota_{\widetilde{k}} \hat{v} = 1$. From \eqref{eq:Ftilde_def} and \eqref{eq:vhat}, we have
\begin{equation}\label{eq:Ftilde_xitilde}
    \widetilde{F} = \dd\widetilde{\xi} \,.
\end{equation}

\subsection{Avoiding the anomaly}

While the isometry of $\hat{M}$ generated by $k$ always gives rise to a $\U(1)$ symmetry on $\hat{M}$, it is not guaranteed to have no 't Hooft anomaly since $\iota_k \hat{v} = \iota_k v$ may be non-zero. However, since the new coordinate $\hat{X}$ does not appear in the metric $\hat{g}$, there is a family of Killing vectors
\begin{align}\label{eq:lift_killing}
    \hat{k} = k + \Theta \partial_{\hat{X}}\,.
\end{align}
For any $\Theta$, we have $\mathcal{L}_{\hat{k}} \hat{g} = \mathcal{L}_{\hat{k}} \hat{H}= 0$ and $\iota_{\hat{k}} \hat{H} = \dd\hat{v}/2\pi$, so the transformation
\begin{equation}\label{eq:symm_transf_Mhat} 
    \delta \hat{X}^I = \alpha \hat{k}^I 
\end{equation}
is a $\U(1)$ symmetry of the NLSM on $\hat{M}$. Furthermore, it has no 't Hooft anomaly if
\begin{equation}
    \mathcal{L}_{\hat{k}} \hat{v} = 0 \qc \iota_{\hat{k}} \hat{v} = 0 \,,
\end{equation}
which is the case if we choose
\begin{equation}\label{eq:Theta=-iv}
    \Theta = -\iota_k v \,.
\end{equation}
That is, the NLSM on $\hat{M}$ always has a non-anomalous $\U(1)$ isometry symmetry associated with the Killing vector \eqref{eq:lift_killing}, with $\Theta$ given by \eqref{eq:Theta=-iv}.

Gauging this symmetry is achieved by the analogue of \eqref{eq:gauged_NLSM_action}:
\begin{align}
\begin{split}\label{eq:gauged_NLSM_action_Mhat}
    \hat{S}_{\text{gauged}} &= \frac{1}{2}\int_W g_{ij}\, DX^i\wedge\star DX^j+\frac{1}{3!}\int_V H_{ijk} \, DX^i\wedge DX^j\wedge DX^k \\
    &\qquad +\frac{1}{2\pi}\int_V dC\wedge \hat{v}_I D\hat{X}^I\,,
\end{split}
\end{align}
where
\begin{equation}
    D\hat{X}^I = d\hat{X}^I - C \hat{k}^I \,.
\end{equation}
Notably, expanding the final term in \eqref{eq:gauged_NLSM_action_Mhat} includes a term
\begin{equation}\label{eq:lag_mult}
    \frac{1}{2\pi} \int_W C \wedge d\hat{X} \,,
\end{equation}
which is the only term in \eqref{eq:gauged_NLSM_action_Mhat} containing the coordinate $\hat{X}$. In other words, $\hat{X}$ appears as a Lagrange multiplier field and integrating it out sets $C$ to a trivial connection (i.e.~flat with no non-trivial holonomies). Then gauge fixing $C=0$ reduces \eqref{eq:gauged_NLSM_action_Mhat} to the action of the original NLSM \eqref{eq:NLSM_action}. 

We now briefly describe this mechanism from the perspective of the original NLSM on $M$, in the case that it has an anomalous $\U(1)$ isometry symmetry. From the perspective of the original NLSM on $M$, the new coordinate $\hat{X}$ introduced above is simply an extra scalar field. The transformation \eqref{eq:symm_transf_Mhat} can be written
\begin{equation}
    \delta X^i = \alpha k^i \qc \delta \hat{X} = \alpha \Theta \,.
\end{equation}
That is, the original scalars $X^i$ transform as they did previously in \eqref{eq:symm_transf}, but the new scalar $\hat{X}$ also transforms under the symmetry. In particular, the term \eqref{eq:lag_mult} which involves $\hat{X}$ transforms as
\begin{equation}
    \frac{1}{2\pi} \int_W C \wedge d\hat{X} \to \frac{1}{2\pi} \int_W C \wedge d\hat{X} - \frac{\iota_k v}{2\pi} \int_W \alpha\, dC \,,
\end{equation}
where we have used \eqref{eq:Theta=-iv} and integrated by parts. From \eqref{eq:Z[C]_variation2}, this precisely cancels the anomalous transformation of the original scalars $X^i$ under the symmetry. Therefore, not only does the extra field $\hat{X}$ in the NLSM on $\hat{M}$ act as a Lagrange multiplier, its transformation is chosen to absorb the anomaly of the original fields $X^i$. In some sense, its role here is similar to that of an axion (albeit without a kinetic term). From this perspective, it is not surprising that the symmetry can always be gauged on $\hat{M}$, since we can always add an axionic field to an anomalous theory to cancel the anomaly.

\subsection{T-duality}
\label{subsec:T_duality}

The target space $\hat{M}$ has two isometries: $\hat{k}$ in \eqref{eq:lift_killing} and $\widetilde{k}$ in \eqref{eq:ktilde}. A more symmetric choice of coordinates on $\hat{M}$ is $( Y^\mu, X, \widetilde{X} )$ where $Y^\mu$ are coordinates on $N$ and \cite{Hull2006}
\begin{equation}
    \hat{k} = \pdv{X} \qc \widetilde{k} = \pdv{\widetilde{X}} \,.
\end{equation}
Above, we first integrated out $\hat{X}$ and gauge fixed $C=0$ to return to the original presentation of the NLSM. Instead, we can also integrate out $C$ since its path integral is Gaussian. Doing so results in a NLSM depending only on the $Y^\mu$ and $\widetilde{X}$ coordinates. The target space $M'$ of this dual model is another $S^1$ bundle \cite{Hull2006}
\begin{align}\label{eq:M'_bundle}
    S^1\hookrightarrow M'\twoheadrightarrow N\,,
\end{align}
with metric
\begin{equation}\label{eq:g'}
    g' = \bar{g} + \frac{1}{(2\pi)^2 G} \widetilde{\xi} \otimes \widetilde{\xi} 
\end{equation}
and flux
\begin{equation}\label{eq:H'}
    H' = \bar{H} + \frac{1}{2\pi} F \wedge \widetilde{\xi} \,.
\end{equation}
We will denote the coordinates on $M'$ by $X^{\prime i} = (Y^\mu, \widetilde{X})$.
This dual geometry has a Killing vector $\widetilde{k}$.
Comparing with \eqref{eq:metric_bundle} and \eqref{eq:H_split}, we see that the roles of the two curvatures $F$ and $\widetilde{F}$ are interchanged in the dual theory. That is, the duality is cleanly described by interchanging
\begin{equation}
    \xi \leftrightarrow \widetilde{\xi} \,,
\end{equation}
which, from \eqref{eq:F_def} and \eqref{eq:Ftilde_xitilde}, implies
\begin{align}\label{eq:F_Ftilde_swap}
    F\leftrightarrow\widetilde{F}\,.
\end{align}
The duality also inverts the radius of the $S^1$ fibre (i.e.~$G\leftrightarrow G^{-1}/(2\pi)^2$), but this will not be relevant for the purposes of this work.

\subsection{Discrete gauging \& duality defects}
\label{subsec:duality_defects}

From \eqref{eq:F_Ftilde_swap}, it is clear that any NLSM described by two bundles with equal curvatures (i.e.~$F=\widetilde{F}$) will be self-T-dual (the theory must also be at the self-T-dual value of $G=1/2\pi$). A simple example of this type is the $SU(2)_1$ WZW model. 

In \cite{Arias-Tamargo:2025xdd}, the doubled torus construction described above was adapted to gauge a $\Z_p$ subgroup of the isometry symmetry. Essentially, the construction allows for a factor of $p$ to be included in the Lagrange multiplier term \eqref{eq:lag_mult}, such that integrating out $\hat{X}$ restricts $C$ to be a flat $\U(1)$ gauge field with $\Z_p$-valued holonomies instead of imposing that $C$ is completely trivial. In other words, this modification sets $C$ to be a $\Z_p$ gauge field and integrating it out then describes the gauging of a $\Z_p$ subgroup of the isometry symmetry. Since $C$ is not a trivial gauge field, this gauging is a non-trivial operation and does not generally describe a duality between the original and gauged theories. The original T-duality construction described in the previous subsection is recovered by setting $p=1$.

For $p\neq1$, while this discrete gauging procedure no longer describes a duality in general, it is possible to find NLSMs which are self-dual under the gauging. The main result of \cite{Arias-Tamargo:2025xdd} is that the conditions for such a self-duality to hold are
\begin{equation}\label{eq:duality_condition}
    \widetilde{F} = pF \qc G = \frac{p}{2\pi} \,.
\end{equation}
The former is a constraint on the topology of the pair of $\U(1)$ bundles describing the NLSM and the latter is a constraint on the parameters of the theory. When these constraints are satisfied, the theory has a non-invertible defect that can be constructed via half-space gauging \cite{ChoiCordovaHsinLam2021}. There are many examples of theories satisfying these constraints, such as the compact boson with $R^2=p/2\pi$ and the $SU(2)_p$ WZW model. The latter is equivalent to the $S^3$ example studied in subsection~\ref{subsec:U(1)isometry} with $K=p$ and $R^2=p/2\pi$.

\section{Dual isometry symmetry}
\label{sec:dual_isom}

So far we have focused on the isometry symmetry of the NLSM. In this section, we discuss another group-like symmetry of the theory. It is not associated with an explicit transformation of the fields and, 
when it is continuous,
its current is identically conserved as a consequence of the Bianchi identity. In the case of the compact boson, it is the familiar winding symmetry, with current $\widetilde{j}=\star dX/(2\pi)$.

\subsection{Dual isometries vs. winding}
\label{subsec:dual_isom_vs_winding}

For more general NLSMs, the situation is not as simple
as for the compact boson.
Let us begin with the example of the SU(2)$_K$ WZW model. This can be described as a NLSM with target space $S^3$ with $K$ units of H-flux and a particular value of the radius $R$ \cite{Witten:1983ar}. In the language of section~\ref{sec:NLSMs}, the target space can be described as an $S^1$ bundle over $S^2$ with Chern number 1, and the H-class associated with the Wess-Zumino term is $K$. The NLSM with this target space was studied in an example in section~\ref{subsec:U(1)isometry}, where it was seen that there is a $\U(1)$ isometry symmetry. However, since $\pi_1(S^3)=0$, there is no winding around the $S^3$ target and the current $\star d\phi$ does not lead to a conserved charge (see e.g.~\cite{Arias-Tamargo:2025xdd} for a recent discussion). 
Intuition from the compact boson would suggest that the isometry and winding symmetries should be exchanged under T-duality so, in particular, the T-dual would be expected to have a $\U(1)$ winding symmetry.
The T-dual model, as per \eqref{eq:F_Ftilde_swap}, is a NLSM whose target space is a Lens space $S^3/\Z_K$ (i.e.~a $S^1$ bundle over $S^2$ with Chern number $K$) with a WZ term with H-class equal to 1. This presents a puzzle as $\pi_1(S^3/\Z_K)=\Z_K$ so the na\"{i}ve winding symmetry would be $\Z_K$, rather than $\U(1)$.
The puzzle was addressed in \cite[Sec. 6]{Maldacena:2001ky}, where it was found that in the Lens space model there is also a U(1) conserved current, although it has no clear geometrical interpretation as a `winding' symmetry; 
that is, it is not associated with $\pi_1(M)$.

This example highlights two important points. Firstly, the fundamental group $\pi_1(M)$ does not (in general) correspond to the symmetry which is dual to the isometry symmetry. Secondly, in a given duality frame some of the global symmetries may not be presented in an obvious manner. Therefore, in order to determine the full global symmetry of the NLSM it is useful to study \emph{both} T-dual descriptions of the theory.

Using the doubled torus construction, the situation is clearer. On $\hat{M}$, there are always two Killing vectors associated with the two $S^1$ fibres, which we have denoted as $\hat{k}$ and $\widetilde{k}$ in section~\ref{sec:T-duality}. The NLSM on $\hat{M}$ can be reduced to either of the two T-dual models on $M$ or $M'$ by gauge-fixing different coordinates. It is only upon reduction to one of the two T-dual NLSMs that the clear geometrical interpretation of each of these two symmetries may be lost. For example, gauge-fixing $\hat{X}$ sends the doubled NLSM on $\hat{M}$ to the original one on $M$. In this presentation of the theory, the symmetry associated with the Killing vector $\hat{k}$ is clear since $\hat{k}$ also generates an isometry of $M$. However, the symmetry associated with $\widetilde{k}$ is less obvious in this presentation of the theory since it is not related to an isometry of $M$.
With this in mind, we refer to the two symmetries related to $\hat{k}$ and $\widetilde{k}$ as the `isometry' and `dual isometry' symmetries respectively in order to avoid confusion with the more standard notion of `winding' symmetries associated with the fundamental group. As seen above in the Lens space example, the symmetry associated with the fundamental group of the target space is generically only a subgroup of the full global symmetry of these NLSMs.

The question is then how to identify the possibly non-geometric dual isometry symmetry of the NLSM on $M$ from the data that specifies the theory. One way to address this is to use T-duality, as discussed in section~\ref{subsec:T_duality}.
We have seen in the previous section that the isometry symmetry is governed by whether the curvature $\widetilde{F}$ is exact in $M$ (which is a bundle described by $F$). T-duality exchanges $F$ and $\widetilde{F}$, so the isometry symmetry of the T-dual NLSM will be governed by whether $F$ is exact or not on the geometry $M'$ in \eqref{eq:M'_bundle} which is a bundle described by $\widetilde{F}$. This is the dual isometry symmetry of the original NLSM. The doubled torus construction suggests that both symmetries (and both bundles) should be considered on the same footing.
Let us now discuss this in more detail.

Let us begin by collecting the relevant formulae for the isometry symmetry and its 't Hooft anomaly. In the original presentation of the theory, $F$ determines the geometry of the target space $M$ in \eqref{eq:M_bundle} (e.g.~its metric is \eqref{eq:metric_bundle}) and $\widetilde{F}$ determines the flux $H$ via \eqref{eq:H_split}. The isometry symmetry associated with the Killing vector $k$ is determined by whether $\widetilde{F}$ is exact or not on $M$. 
The nature of the symmetry depends on the value of $\kappa$ in \eqref{eq:kappa_gcd}, which can be written
\begin{equation}\label{eq:kappa_repeat}
    \kappa = \abs{\gcd( \kappa_i )}\qc \kappa_i = \frac{1}{2\pi} \int_{\Sigma_i} \widetilde{F} \,,
\end{equation}
where $\{\Sigma_i\}$ is a basis of $H_2(M)$. There is a $\U(1)$ symmetry if $\kappa=0$, which is the case if and only if all the $\kappa_i=0$ and $\widetilde{F}$ is exact on $M$. If $\kappa\neq0$ then there is a $\Z_\kappa$ isometry symmetry.
When there is a $\U(1)$ isometry symmetry, we can write $\widetilde{F}=\dd v$ for a globally defined 1-form $v \in \Omega^1(M)$ and the isometry symmetry has a 't Hooft anomaly if the bundle \eqref{eq:M_bundle} described by $F$ is non-trivial and $\iota_k v \neq0$. There is a non-anomalous $\Z_p$ subgroup where $p$ is given by \eqref{eq:p_def}, which we repeat here for convenience:
\begin{equation}\label{eq:p}
    p=\abs{\iota_k v} \,.
\end{equation}
If the basic 2-form $F$ is exact on $N$ then $v$ can be chosen such that $\iota_k v=0$ and there is no anomaly.
When there is only a discrete $\Z_\kappa$ isometry symmetry, there is no 't Hooft anomaly either.

We now consider the T-dual frame and apply the same logic. Everything is the same, but the roles of $F$ and $\widetilde{F}$ are reversed. Namely, $\widetilde{F}$ determines the geometry of the target space $M'$ in \eqref{eq:M'_bundle} (with metric \eqref{eq:g'}) and $F$ determines the flux $H'$ via \eqref{eq:H'}. The Killing vector $\widetilde{k}$ gives rise to a dual isometry symmetry. The symmetry group depends on the value of $\widetilde{\kappa}$ given by
\begin{equation}\label{eq:kappatilde}
    \widetilde{\kappa} = \abs{\gcd(\widetilde{\kappa}_i)}\qc 
    \widetilde{\kappa}_i = \frac{1}{2\pi} \int_{\widetilde{\Sigma}_i} F \,,
\end{equation}
where $\{\widetilde{\Sigma}_i\}$ is a basis of $H_2(M')$. There is a $\U(1)$ symmetry if $\widetilde{\kappa}=0$, which is the case if and only if all the $\widetilde{\kappa}_i=0$ and $F$ is exact on $M'$.
If $\widetilde{\kappa}\neq0$, there is a $\Z_{\widetilde{\kappa}}$ symmetry.
If the dual isometry symmetry is $\U(1)$, we have $F=\dd\widetilde{v}$ for a globally-defined 1-form $\widetilde{v}\in \Omega^1(M')$ and the dual isometry symmetry has a 't Hooft anomaly if the bundle \eqref{eq:M'_bundle} described by $\widetilde{F}$ is non-trivial and $\iota_{\widetilde{k}} \widetilde{v} \neq0$. There is a non-anomalous $\Z_{\widetilde{p}}$ subgroup where 
\begin{equation}\label{eq:ptilde}
    \widetilde{p}=\abs{\iota_{\widetilde{k}} \widetilde{v}} \,.
\end{equation}
If the basic 2-form $\widetilde{F}$ is exact on $N$ then $\widetilde{v}$ can be chosen such that $\iota_{\widetilde{k}}\widetilde{v}=0$ and there is no 't Hooft anomaly.
When there is a discrete $\Z_{\widetilde{\kappa}}$ dual isometry symmetry, there is also no anomaly.

A summary of the properties of the isometry and dual isometry symmetries is given in Table~\ref{tab:symmetries}. 

\begin{table}[t]
    \centering
    \begin{tabular}{c c c}
        \hline\\[-2.5ex]
        \textbf{NLSM property} & \textbf{Symmetry} & \textbf{Anomaly} \\ [0.5ex] 
        \hline\hline \\[-2ex] 
        $\widetilde{F}$ exact on $M$ & $\U(1)$ isometry & $p=\abs{\iota_k v}$ \\
        $\widetilde{F}$ not exact on $M$ & $\Z_\kappa$ isometry & -- \\
        $F$ exact on $M'$ & $\U(1)$ dual isometry & $\widetilde{p} = \abs{\iota_{\widetilde{k}} \widetilde{v}}$ \\
        $F$ not exact on $M'$ & $\Z_{\widetilde{\kappa}}$ dual isometry & --\\[1ex]
        \hline
    \end{tabular}
    \caption{The isometry and dual isometry symmetries in the NLSM are dictated by properties of $F$ and $\widetilde{F}$. In the discrete cases, $\kappa$ and $\widetilde{\kappa}$ are given by \eqref{eq:kappa_repeat} and \eqref{eq:kappatilde} respectively.}
    \label{tab:symmetries}
\end{table}

It is clear from the above discussion that the isometry and dual isometry symmetries are interchanged by T-duality. This can also be seen from the currents associated with the symmetries in a simple case.
We consider the case where both the isometry and dual isometry symmetries are non-anomalous $\U(1)$ symmetries. This is the case when $F$ and $\widetilde{F}$ are exact 2-forms on the base $N$.
Recall from \eqref{eq:isom_current} that the Noether current for the isometry symmetry is
\begin{equation}\label{eq:isom_current_repeat}
    j = g_{ij} k^j dX^i - \frac{1}{2\pi} v_i \star dX^i \,.
\end{equation}
Similarly, in the other duality frame, the Noether current $\widetilde{j}$ for the dual isometry symmetry is
\begin{equation}\label{eq:dual_isom_current}
    \widetilde{j} = g'_{ij} \widetilde{k}^{j} dX^{\prime i} - \frac{1}{2\pi} \widetilde{v}_{i} \star dX^{\prime i} \,.
\end{equation}
However, the latter is written in terms of the coordinates $X^{\prime i} = (Y^\mu, \widetilde{X})$ on $M'$.
In order to describe both $\U(1)$ symmetries in the original duality frame (i.e.~in terms of coordinates $X^i$ on $M$) we need to understand how these currents are mapped under T-duality. This has been discussed recently in \cite{Arias-Tamargo:2025xdd} (and references therein), with the result being that the dual isometry current $\widetilde{j}$ can be written
\begin{equation}
    \widetilde{j} = \frac{1}{2\pi} \star dX
\end{equation}
in the original duality frame. Indeed, this current is identically conserved, $d\star \widetilde{j}=0$. This current is a well-defined operator since $dX$ is a globally defined 1-form on $M$. This is because $F$ is exact on $N$ in this example,
which implies that the bundle \eqref{eq:M_bundle} describing $M$ as a $\U(1)$ bundle over $N$ is trivial and the 1-form $dX$ along the fibre is a globally defined closed 1-form.
If either $F$ is not exact on $M'$ or $\widetilde{F}$ is not exact on $M$ then one (or both) of the $\U(1)$ symmetries is broken to a discrete subgroup and the symmetry cannot be described in terms of the Noether currents.

\subsection{Mixed anomaly of isometry and dual isometry symmetries}
\label{subsec:mixed_anom}

There are many examples of theories with a pair of global symmetries related by a duality which have a mixed 't Hooft anomaly. For example, this is common in free theories such as the compact boson, Maxwell theory \cite{Gaiotto:2014kfa}, and linear gravity \cite{Hull:2024bcl, Hull:2024ism}. The situation here is no different; that is, the isometry and dual isometry symmetries participate in a mixed 't Hooft anomaly, as we now show.

Let us begin with the case where there is both a $\U(1)$ isometry symmetry and a $\U(1)$ dual isometry symmetry, both of which have no pure 't Hooft anomaly. As discussed above, in this case the Noether currents for the two symmetries can be written in a single duality frame as \eqref{eq:isom_current_repeat} and \eqref{eq:dual_isom_current}. In order to diagnose a mixed 't Hooft anomaly, we must couple background $\U(1)$ gauge fields to both symmetries and see whether the path integral is invariant under background transformations of both. Coupling the background field $C$ for the $\U(1)$ isometry symmetry was discussed in subsection~\ref{subsec:anomalies_U(1)}. In the case where the isometry symmetry has no pure 't Hooft anomaly (i.e.~a global choice of $v$ exists such that $\iota_k v=0$), the gauged action is given by $S^W_{\text{gauged}}$ in \eqref{eq:S_W_gauged}. 

Now consider gauging the dual isometry symmetry. In the duality frame with target space $M$, it does not have a local action on the fields. This is guaranteed by the fact that the current $\widetilde{j}$ is identically conserved. It can be gauged simply by adding a linear term coupling a $\U(1)$ background gauge field $\widetilde{C}$ to the current $\widetilde{j}$,
\begin{equation}
    S + \frac{1}{2\pi} \int_W \widetilde{C} \wedge dX \,.
\end{equation}
This is invariant under background gauge transformations $\widetilde{C} \to \widetilde{C} + d\widetilde{\alpha}$. 

We now try to introduce couplings to both background fields simultaneously. The obvious starting point is simply to combine the two gaugings discussed above for each symmetry individually. Namely, we consider an action
\begin{equation}\label{eq:Sboth}
    S_{\text{both}} = S^W_{\text{gauged}} + \frac{1}{2\pi} \int_W \widetilde{C} \wedge dX \,.
\end{equation}
While this is invariant under $\widetilde{C} \to \widetilde{C} + d\widetilde{\alpha}$ (which does not act on the $S^W_{\text{gauged}}$ term), it is not invariant under the isometry symmetry transformation: $X \to X+\alpha$, $C \to C + d\alpha$. In particular, the final term in \eqref{eq:Sboth} is not invariant and we find
\begin{equation}\label{eq:Sboth_variation}
    S_{\text{both}} \to S_{\text{both}} + \frac{1}{2\pi} \int_W \widetilde{C} \wedge d\alpha \,.
\end{equation}
Before concluding that there is a mixed 't Hooft anomaly, we must check that there is no local two-dimensional counter-term involving the background fields which can be added to remove this lack of invariance. One can, for example, add a term $\int_W C \wedge \widetilde{C} /2\pi$. The resulting action is invariant under the isometry symmetry transformations, but it then fails to be invariant under the dual isometry transformation.

It is straightforward to verify that it is not possible to find local two-dimensional couplings to both background fields which are simultaneously gauge-invariant under both transformations, indicating a mixed 't Hooft anomaly. The lack of invariance can be described by an anomaly inflow theory with action
\begin{equation}\label{eq:S_inflow_mixed}
    S_{\text{inflow}}^{\text{mixed}} = -\frac{1}{2\pi} \int_V C \wedge d\widetilde{C} \,,
\end{equation}
whose variation under the isometry symmetry transformation precisely cancels that of $S_{\text{both}}$ in \eqref{eq:Sboth_variation}, so $S_{\text{both}} + S_{\text{inflow}}^{\text{mixed}}$ is invariant. 
While the inflow theory \eqref{eq:S_inflow_mixed} is well-known in the case of the compact boson, interestingly we see that the same mixed anomaly is present for all NLSMs with these symmetries.

Here we have considered the case where both the isometry and dual isometry symmetries have no pure 't Hooft anomalies in order to focus on the mixed 't Hooft anomaly. In the most general case, these two symmetries can also have pure 't Hooft anomalies and the total inflow action which describes the individual and mixed anomalies is
\begin{equation}\label{eq:inflow_total}
    S_{\text{inflow}}^{\text{total}} = - \frac{p}{2\pi} \int_V C\wedge dC - \frac{\widetilde{p}}{2\pi} \int_V \widetilde{C} \wedge d\widetilde{C} - \frac{1}{2\pi} \int_V C \wedge d\widetilde{C} \,,
\end{equation}
where $p$ and $\widetilde{p}$ are given by \eqref{eq:p} and \eqref{eq:ptilde} respectively.
The results of Appendix~\ref{app:quantisation} imply that the prefactors of each term are quantised appropriately such that this inflow action gives a well-defined theory. Namely, $p$ and $\widetilde{p}$ are integers.

\subsection{Mixed anomaly in the discrete case}

In the next section, we will be interested in gauging a discrete subgroup of the isometry symmetry. Of course, in order for this gauging to be possible, this discrete subgroup must be non-anomalous. From \eqref{eq:inflow_total}, there is a non-anomalous $\Z_p$ subgroup of the isometry symmetry.\footnote{Any subgroup of this $\Z_p$ is also non-anomalous and the manipulations below can be straightforwardly modified to consider any such subgroup instead of the full $\Z_p$.} In particular, we will be interested in the mixed 't Hooft anomaly between this $\Z_p$ subgroup of the isometry symmetry and the dual isometry symmetry. If there was only a discrete $\Z_\kappa$ isometry symmetry to begin with, we know from subsection~\ref{subsec:anomalies_discrete} that there is no pure 't Hooft anomaly and so all subgroups are non-anomalous (i.e.~$p$ can be any divisor of $\kappa$ in this case).

Restricting the $\U(1)$ isometry symmetry background field $C$ to a $\Z_p$ subgroup is achieved by replacing
\begin{equation}
    C \to \frac{2\pi}{p} C_p\,,
\end{equation}
where $C_p \in H^1(V,\Z_p)$ is a discrete gauge field. This relation should be understood as restricting the holonomies $\exp$$(i\int C)$ of the $\U(1)$ connection $C$ to lie in a $\Z_p$ subgroup, see Appendix~\ref{app:bockstein} for more details. The reduction of the mixed anomaly inflow action \eqref{eq:S_inflow_mixed} is
\begin{equation}\label{eq:mixed_anom_discrete}
    S_{\text{inflow}}^{\text{discrete}} = \frac{2\pi}{p} \int_V C_p \cup \left[ \frac{d\widetilde{C}}{2\pi} \text{ mod } p \right] \,,
\end{equation}
where $\cup$ is the standard cup product of $\Z_p$ cochains and $[\cdots]$ denotes the cohomology class in $H^2(V,\Z_p)$. Consider now a $\Z_q$ subgroup of the $\U(1)$ dual isometry symmetry. The background field $\widetilde{C}$ can be restricted to this subgroup via 
\begin{equation}\label{eq:int_lift_relation}
    \widetilde{C} \to \frac{2\pi}{q} \widetilde{C}_q
\end{equation}
where $\widetilde{C}_q \in H^1(V,\Z_q)$. 
The $\U(1)$ background field $\widetilde{C}$ can be understood as an integral lift of the discrete field $\widetilde{C}_q$ (see Appendix~\ref{app:bockstein}). The cohomology class appearing in \eqref{eq:mixed_anom_discrete} is an expression for the Bockstein homomorphism associated with the short exact sequence
\begin{equation}\label{eq:Zp_SES}
    1 \to \Z_p \to \Z_{pq} \to \Z_q \to 1 \,.
\end{equation}
In other words, we have
\begin{equation}\label{eq:int_lift_bock}
    \left[ \frac{d\widetilde{C}}{2\pi} \text{ mod } p \right] = \Bock([\widetilde{C}_q]) \,.
\end{equation}
Intuitively, the Bockstein plays the role of the curvature of the discrete gauge field $\widetilde{C}_q$.
Using \eqref{eq:int_lift_bock}, the mixed anomaly \eqref{eq:mixed_anom_discrete} can then be written
\begin{equation}\label{eq:mixed_anom_bock}
    S_{\text{inflow}}^{\text{discrete}} = \frac{2\pi}{p} \int_V C_p \cup \Bock([\widetilde{C}_q]) \,.
\end{equation}
This expression is of particular relevance if the isometry and dual isometry symmetries of the NLSM are discrete rather than $\U(1)$. The isometry and dual isometry symmetries are only $\U(1)$ if either $F$ or $\widetilde{F}$ is exact, which is non-generic. From subsection~\ref{subsec:dual_isom_vs_winding}, generically there is a $\Z_\kappa$ isometry symmetry and a $\Z_{\widetilde{\kappa}}$ dual isometry symmetry with $\kappa$ and $\widetilde{\kappa}$ given by \eqref{eq:kappa_repeat} and \eqref{eq:kappatilde} respectively.

Since only the cohomology class $\Bock(\widetilde{C}_q)$ is relevant to the anomaly \eqref{eq:mixed_anom_bock}, the anomaly vanishes whenever this class is trivial. In particular, this happens when the short exact sequence \eqref{eq:Zp_SES} splits, which is the case when $\gcd(p,q)=1$ (see Appendix~\ref{app:bockstein}). Indeed, this is necessary of a mixed $\Z_p$-$\Z_q$ anomaly. One way to see this, following \cite{Csaki:1997aw}, is because the charges of a $\Z_p$ symmetry are only defined modulo $p$, so the anomaly coefficient of a mixed $\Z_p$-$\Z_q$ anomaly should be insensitive to any shift of the form $ap+bq$ where $a,b\in\Z$. By Bézout's lemma, the integers $ap+bq$ are the multiples of $\gcd(p,q)$ so the mixed anomaly is therefore insensitive to a shift of the anomaly coefficient by any multiple of $\gcd(p,q)$. In other words, the anomaly coefficient is defined modulo $\gcd(p,q)$. Therefore, if $\gcd(p,q)=1$ the anomaly must be trivial.

\section{Discrete gauging \& symmetry structure}\label{sec:extensions}

Having understood the global group-like symmetries of the NLSM and their anomalies, in this section we study how these symmetry structures behave under the gauging of a non-anomalous discrete subgroup of the isometry symmetry. This question has been analysed in detail in \cite{Tachikawa:2017gyf}, and we will use the results of that work heavily. Furthermore, in NLSMs admitting a non-invertible duality defect (such as those studied in \cite{Arias-Tamargo:2025xdd}), the discrete gauging maps the original theory back to itself and so the symmetry structure must be invariant. We will show that when the NLSM satisfies the self-duality conditions found in \cite{Arias-Tamargo:2025xdd}, the symmetry structure is indeed preserved by the gauging.

We know that a NLSM with a target space $M$ which has a freely acting isometry generically has both an isometry and a dual isometry symmetry. The isometry symmetry is generically $\Z_\kappa$ where $\kappa$ is an integer given in \eqref{eq:kappa_repeat}. From the discussion in subsection~\ref{subsec:anomalies_discrete}, this discrete symmetry does not have a 't Hooft anomaly. The situation is slightly different when $\kappa=0$. In this case there is instead a $\U(1)$ isometry symmetry which generically has an anomaly \eqref{eq:pure_inflow}. From section~\ref{subsec:anomalies_U(1)}, there is a non-anomalous $\Z_p$ subgroup with $p=\abs{\iota_k v}$. The dual isometry symmetry has an analogous structure and is generically $\Z_{\widetilde{\kappa}}$ where $\widetilde{\kappa}$ is an integer given by \eqref{eq:kappatilde}. This discrete symmetry does not have a 't Hooft anomaly. When $\widetilde{\kappa}=0$ there is instead a $\U(1)$ dual isometry symmetry which generically has an anomaly, but there is a non-anomalous $\Z_{\tilde{p}}$ subgroup with $\widetilde{p} = \abs{\iota_{\widetilde{k}} \widetilde{v}}$.

\subsection{The exact sequence}

Let us focus on the isometry symmetry in the case where $\kappa\neq0$, so there is a $\Z_\kappa$ isometry symmetry. Suppose $p$ is a divisor of $\kappa$, then $\Z_p$ is a non-anomalous normal subgroup and there is a short exact sequence
\begin{equation}\label{eq:discrete_SES} 
    1 \to \Z_p \xrightarrow{\iota} \Z_\kappa \xrightarrow{\pi} \Z_r \to 1
\end{equation}
where $r=\kappa/p$.

In the case where $\kappa=0$ and there is a $\U(1)$ isometry symmetry, there is a non-anomalous $\Z_p$ subgroup with $p$ given by \eqref{eq:p} and a similar sequence
\begin{equation}
    1 \to \Z_p \to \U(1) \to \U(1) \to 1\,.
\end{equation}
In either case, there is a non-anomalous $\Z_p$ subgroup which can be gauged. The only difference is that in the discrete case $p$ can be any divisor of $\kappa$, while in the continuous case it must take the value \eqref{eq:p}, or a divisor thereof. In the following we will focus on the discrete case (i.e.~when $\kappa\neq0$), but the results can be simply extended to include the $\kappa=0$ case.

\subsection{Gauging discrete subgroups}

We are interested in gauging the non-anomalous $\Z_p$ subgroup of the $\Z_\kappa$ isometry symmetry. The gauged theory will have a residual $\Z_r = \Z_\kappa/\Z_p$ global symmetry as well as a dual $\widehat{\Z}_p=\Z_p$ symmetry. The dual symmetry is generated by codimension-1 topological defects which are the Wilson lines for the dynamical $\Z_p$ gauge field in the gauged theory.
In the context of two-dimensional conformal field theories (CFTs), this is often referred to as the quantum symmetry of $\Z_p$ under which the twisted sectors of the ungauged theory carry charge. If the sequence \eqref{eq:discrete_SES} does not split, these two global symmetries in the gauged theory have a mixed 't Hooft anomaly \cite{Tachikawa:2017gyf}, as we now briefly review.

Let $A_\kappa$ be a background field for the $\Z_\kappa$ global symmetry; that is, $A_\kappa$ is a 1-cocycle representing a class $[A_\kappa] \in H^1(W,\Z_\kappa)$. From the short exact sequence \eqref{eq:discrete_SES}, we can write
\begin{equation}\label{eq:field_split}
	A_\kappa = \iota(A_p) + A'_r \,,
\end{equation}
where $A'_r \in C^1(W,\Z_\kappa)$ is an arbitrary lift of the 1-cocycle $A_r=\pi(A_\kappa)$ representing a class $[A_r] \in H^1(W,\Z_r)$, so $\pi(A'_r) = A_r$. Since $\pi$ is surjective such a lift exists, but generically it is not a cocycle. As $\pi(A_\kappa-A'_r)=0$, the exactness of the sequence implies that $A_\kappa$ can be written as in \eqref{eq:field_split} with $A_p \in C^1(W,\Z_p)$. 

Now, from $\delta A_\kappa =0$ we have $\iota(\delta A_p)+\delta A_r'=0$. Using that $\pi(\delta A_r') = \delta A_r = 0$ and the exactness of the sequence \eqref{eq:discrete_SES}, we find that there exists a cocycle $\nu_r \in C^2(W,\Z_p)$ such that $\delta A_r' = \iota(\nu_r)$. The class represented by $\nu_r$ is denoted $[\nu_r] = \Bock([A_r])$, where $\Bock:H^1(W,\Z_r) \to H^2(W,\Z_p)$ is the Bockstein homomorphism associated with the short exact sequence \eqref{eq:discrete_SES} (see Appendix~\ref{app:bockstein}). Now since $\iota$ is injective, we find
\begin{equation}\label{eq:da=Bock}
	\delta A_p = - \nu_r \,.
\end{equation}
That is, the $\Z_p$ background field may not be a cocycle due to the non-trivial short exact sequence.\footnote{This structure is similar in spirit to the Green-Schwarz mechanism, where the closure of the 3-form $H$ is determined by the curvature of another gauge field as $\dd{H}\sim \tr F\wedge F$. A similar structure is used to describe background fields for 2-group symmetries in \cite{Kapustin:2013uxa, Benini:2018, Tachikawa:2017gyf}.} We do note, however, that \eqref{eq:da=Bock} implies $\Bock([A_r])=0$ is a trivial cohomology class. Indeed, this is to be expected since the Bockstein measures the obstruction to lifting a class in $H^1(W,\Z_r)$ to a class in $H^1(W,\Z_\kappa)$. Since $A_r=\pi(A_\kappa)$ and $[A_\kappa] \in H^1(W,\Z_\kappa)$ by definition, such a lift exists and $\Bock([A_r])$ is trivial.\footnote{\label{fn:bock}This does \emph{not} mean that the short exact sequence \eqref{eq:discrete_SES} splits. This is seen explicitly by recalling that the sequence \eqref{eq:discrete_SES} is specified by a group 2-cocycle $e \in H^2(B\Z_r,\Z_p)$ (see Appendix~\ref{app:bockstein}), and the $\Z_r$ gauge field $A_r$ specifies a map, which we will also call $A_r:W\to B\Z_r$. The Bockstein can be understood as the pull-back of the group cocycle along this map: $\Bock([A_r]) = A_r^* e$ \cite{Tachikawa:2017gyf}. Therefore, given the sequence \eqref{eq:discrete_SES}, $\Bock([A_r])=0$ is understood as a constraint on $A_r$, not on $e$ \cite{Bhardwaj:2021wif}. On the other hand, if the sequence \eqref{eq:discrete_SES} splits then $e=0$ and the Bockstein is trivial for arbitrary backgrounds $A_r$.}

Gauging the $\Z_p$ subgroup involves making the background $A_p$ dynamical. We will denote the dynamical field by a lowercase $a_p$. Let us denote by $B_p \in H^1(W,\Z_p)$ a background gauge field for the dual $\widehat{\Z}_p$ symmetry in the gauged theory. Coupling this background in the gauged theory is achieved by introducing a term
\begin{equation}\label{eq:S_dual}
	S_\text{dual} = \frac{2\pi}{p} \int_W B_p \cup a_p
\end{equation}
in the action. This can be understood as inserting a network of $\Z_p$ Wilson lines on a cycle which is Poincaré dual to $B_p$.

Under a background gauge transformation $B_p \to B_p + \delta \epsilon_p$, the variation of the action of the gauged theory is
\begin{equation}\label{eq:dual_variation}
	\Delta S_\text{dual} = \frac{2\pi}{p} \int_W \delta \epsilon_p \cup a_p = - \frac{2\pi}{p} \int_W \epsilon_p \cup \delta a_p = \frac{2\pi}{p} \int_W \epsilon_p \cup \nu_r \,,
\end{equation}
where we have used \eqref{eq:da=Bock} in the final equality. 

This implies that there is a mixed 't Hooft anomaly in the gauged theory between the dual $\Z_p$ symmetry and the residual $\Z_r$ symmetry. In particular, the variation \eqref{eq:dual_variation} is described by the inflow action
\begin{equation}\label{eq:mixed_anom_from_ext}
	S_\text{inflow}^\text{mixed} = \frac{2\pi}{p} \int_V B_p \cup \nu_r \,.
\end{equation}
We see that gauging a normal subgroup of a global symmetry which is involved in a group extension results in a mixed 't Hooft anomaly in the gauged theory.
While $[\nu_r]=\Bock([A_r])=0$, this does not imply that the anomaly vanishes in general. By definition there is no 't Hooft anomaly if a counter-term can be found which cancels the variation of the path integral and is a local expression of the background fields only.
In the present case, in order to find such a counter-term we would need to be able to write $\nu_r = \delta \xi_r$ where $\xi_r \in C^1(W,\Z_p)$ is a cochain given by a local expression of the background fields $A_r, B_p$, since then the inflow \eqref{eq:mixed_anom_from_ext} could be written as a two-dimensional counter-term $(2\pi/p)\int_W B_p \cup \xi_r$. Merely knowing that $[\nu_r]=0$ does not guarantee that $\xi_r$ can be written in this form.\footnote{Relatedly, if the inflow \eqref{eq:mixed_anom_from_ext} were considered on a closed manifold $V$, it would vanish since it would depend only on the cohomology classes of $B_p$ and $\nu_r$, and $[\nu_r]=0$. However, since $\partial V=W$, this is not the case here. Indeed, the variation of \eqref{eq:mixed_anom_from_ext} reproduces \eqref{eq:dual_variation}.}
When the sequence \eqref{eq:discrete_SES} splits, on the other hand, the Bockstein is trivial for arbitrary backgrounds $A_r$, so $\nu_r=0$ and the anomaly vanishes (see footnote~\ref{fn:bock}). 

The inverse is also true: given two symmetries with a mixed 't Hooft anomaly of this type, gauging one of them results in a theory with a global symmetry described by a non-trivial group extension. This can be inferred immediately from the discussion above since gauging the dual $\Z_p$ symmetry gives back the original theory. Therefore, given a theory with a $\Z_p \times \Z_r$ global symmetry with a mixed 't Hooft anomaly of the form \eqref{eq:mixed_anom_from_ext}, gauging the $\Z_p$ symmetry results in a theory with a $\Z_\kappa$ global symmetry described by the group extension \eqref{eq:discrete_SES}.

This can also be seen more explicitly as follows \cite{Benini:2018}. Gauging the $\Z_p$ symmetry implies making the background field $B_p$ dynamical. We will denote the dynamical field by $b_p$. As above, gauging this discrete symmetry results in a dual $\widehat{\Z}_p = \Z_p$ symmetry generated by the Wilson lines of $b_p$. We denote the background gauge field for this dual symmetry by $A_p$. The coupling to $A_p$ in the gauged theory is via a coupling similar to \eqref{eq:S_dual},
\begin{equation}\label{eq:S_dual'}
	S_\text{dual}' = -\frac{2\pi}{p} \int_W b_p \cup A_p \,.
\end{equation}
Since $b_p$ is dynamical, the theory must be invariant under its gauge transformation $b_p \to b_p + \delta \epsilon_p$. The variation of the action of the gauged theory under this transformation has two contributions:
one comes from the mixed 't Hooft anomaly \eqref{eq:mixed_anom_from_ext} of the ungauged theory, and the other from the coupling \eqref{eq:S_dual'}. In total, we must have
\begin{equation}
	0 = \frac{2\pi}{p} \int_W \epsilon_p \cup ( \nu_r + \delta A_p ) 
\end{equation}
to ensure gauge invariance. The background $A_p$ for the dual $\Z_p$ symmetry is, therefore, not closed but instead satisfies precisely the relation found above in \eqref{eq:da=Bock}.
Overall, the symmetries of the gauged theory are described by a $\Z_p$-valued field $A_p$ and a $\Z_r$-valued field $A_r$ satisfying \eqref{eq:da=Bock}.
These can be combined into $A_\kappa$ in \eqref{eq:field_split} which is then a background field for a $\Z_\kappa$ symmetry.
In summary, given a theory with a $\Z_p\times\Z_r$ symmetry with a mixed 't Hooft anomaly \eqref{eq:mixed_anom_from_ext}, gauging the $\Z_p$ symmetry results in a theory with a $\Z_p$ and $\Z_r$ symmetry which combine into a non-trivial group extension \eqref{eq:discrete_SES}. That is, the gauged theory has a $\Z_\kappa$ global symmetry.

\subsection{Self-dualities}

We now apply the results of the previous subsections to the symmetry structure of NLSMs discussed in section~\ref{sec:dual_isom}. We consider the generic situation where the NLSM has a $\Z_\kappa$ isometry symmetry and a $\Z_{\widetilde{\kappa}}$ dual isometry symmetry, with $\kappa$ and $\widetilde{\kappa}$ given by \eqref{eq:kappa_repeat} and \eqref{eq:kappatilde} respectively. The isometry symmetry has a non-anomalous $\Z_p$ subgroup where $p|\kappa$. The full $\Z_\kappa$ isometry symmetry can be thought of as a central extension of this subgroup as described by the short exact sequence \eqref{eq:discrete_SES}, which we repeat here for convenience:
\begin{equation}\label{eq:SES_repeat}
    1 \to \Z_p \xrightarrow{\iota} \Z_\kappa \xrightarrow{\pi} \Z_r \to 1 \,,
\end{equation}
with $r=\kappa/p$. From \eqref{eq:mixed_anom_discrete}, this $\Z_p$ subgroup is involved in a mixed anomaly with the dual isometry symmetry described by the inflow action
\begin{equation}\label{eq:discrete_anomaly_NLSM}
    S_{\text{inflow}}^{\text{discrete}} = \frac{2\pi}{p} \int_V C_p \cup \nu_{\widetilde{\kappa}}\,,
\end{equation}
where $C_p$ is the background field for the $\Z_p$ subgroup of the isometry symmetry (that is, the background field $C_\kappa$ for the full $\Z_\kappa$ isometry symmetry can be decomposed as in \eqref{eq:field_split}) and $\nu_{\widetilde{\kappa}}\in C^2(W,\Z_p)$ is a cocycle representing the class $[\nu_{\widetilde{\kappa}}] = \Bock(\widetilde{C}_{\widetilde{\kappa}})$, where $\widetilde{C}_{\widetilde{\kappa}}$ is the background field for the $\Z_{\widetilde{\kappa}}$ dual isometry symmetry. The Bockstein homomorphism appearing in \eqref{eq:discrete_anomaly_NLSM} is associated with the short exact sequence
\begin{equation}\label{eq:extension_NLSM}
    1 \to \Z_p \to \Z_{p\widetilde{\kappa}} \to \Z_{\widetilde{\kappa}} \to 1 \,.
\end{equation}

We now gauge the $\Z_p$ subgroup. The results of the previous subsection tell us how the symmetry structure is affected. First, the group extension \eqref{eq:SES_repeat} in the ungauged theory implies that in the gauged theory there is a mixed anomaly between the dual $\widehat{\Z}_p$ symmetry and the residual $\Z_r$ symmetry. Second, the mixed anomaly \eqref{eq:discrete_anomaly_NLSM} involving the $\Z_p$ subgroup in the ungauged theory implies that the $\Z_{\widetilde{\kappa}}$ dual isometry symmetry of the ungauged theory is extended by the dual $\widehat{\Z}_p$ symmetry according to the sequence \eqref{eq:extension_NLSM}. That is, the gauged theory has a $\Z_{p\widetilde{\kappa}}$ global symmetry.

In other words, in the original theory the symmetry structure is described by a pair of short exact sequences \eqref{eq:SES_repeat} and \eqref{eq:extension_NLSM}. The former describes the embedding of the non-anomalous $\Z_p$ subgroup of the isometry symmetry and the latter describes the mixed anomaly between this subgroup and the dual isometry symmetry. In the gauged theory, the role of these two sequences is interchanged. 
Clearly then, the symmetry structure of the original and gauged theories are the same if and only if these two sequences are the same; that is, if
\begin{equation}\label{eq:duality_condition_kappa}
    \kappa = p\widetilde{\kappa} \,.
\end{equation}

We recall from subsection~\ref{subsec:duality_defects} that a NLSM is self-dual under gauging a $\Z_p$ subgroup of the isometry symmetry if the conditions \eqref{eq:duality_condition} are satisfied. Focussing in particular on the constraint on the curvatures, $\widetilde{F} = pF$, we see from \eqref{eq:kappa_repeat} and \eqref{eq:kappatilde} that such NLSMs precisely satisfy \eqref{eq:duality_condition_kappa}. That is, we explicitly see that the symmetry structure of the NLSM is preserved by the discrete gauging.

\section{Summary \& outlook}\label{sec:outlook}

In this work, we have discussed in detail the group-like global symmetries of Non-Linear Sigma Models, their anomalies, and discussed their role in the recent construction of non-invertible symmetries of these models in \cite{Arias-Tamargo:2025xdd}.

In the first part of the paper, we have discussed the isometry and dual isometry symmetries of NLSMs, which generalise the momentum and winding symmetries of the compact boson. The NLSM is specified by two $\U(1)$ bundles with curvatures $F$ and $\widetilde{F}$. Whether these symmetries are continuous or discrete depends on whether the relevant curvature 2-form is exact or not. For a U(1) isometry symmetry, we need $\widetilde{F}$ to be exact in the target space $S^1\hookrightarrow M\twoheadrightarrow N$, which is a fibration specified by $F$ \cite{Hull:1989jk}. For a U(1) dual isometry symmetry, we need $F$ to be exact in the target space $S^1\hookrightarrow M'\twoheadrightarrow N$, which is a fibration specified by $\widetilde{F}$. If they are closed, but not exact, only discrete subgroups $\Z_\kappa$ and $\Z_{\widetilde{\kappa}}$ will respectively survive as a global symmetry, where $\kappa$ is specified by the H-class in \eqref{eq:kappa_repeat} and $\widetilde{\kappa}$ is specified by the Chern class in \eqref{eq:kappatilde}.

We also studied the 't Hooft anomalies of these symmetries, both in the continuous and the discrete case. 
The anomaly of the isometry (resp. dual isometry) symmetry is determined by whether $\widetilde{F}$ (resp. $F$) is an exact 2-form on the base $N$, in addition to being exact on $M$ (resp. $M'$).
When it is not, the corresponding anomaly inflow theory is a Chern-Simons theory with level given by the component $\iota_k v$ (resp. $\iota_{\widetilde{k}} \widetilde{v}$)
where $v$ and $\widetilde{v}$ are 1-forms that satisfy $\widetilde{F}=\dd v$, $F=\dd \widetilde{v}$.

Even when the pure 't Hooft anomalies are absent, the isometry and dual isometry symmetries have a mixed 't Hooft anomaly, exactly as in the case of the compact boson. This is true whether the symmetries are continuous or discrete. 

These symmetries and their anomalies have an important role in the construction of non-invertible symmetries in NLSMs of \cite{Arias-Tamargo:2025xdd}, which we studied in the last part of the paper. This non-invertible symmetry is found by gauging a finite subgroup of the global symmetry in half of the worldsheet and looking for the particular values of the various parameters such that the theory on both halves is the same \cite{ChoiCordovaHsinLam2021}. Two zeroth order requirements for this to be possible are that 1) there is a non-anomalous discrete subgroup to gauge and 2) the global symmetry is preserved by the gauging. Our findings allow us to conclude that this is the case: there is always a discrete subgroup of the global symmetry that one can gauge and, moreover, the symmetry structure is left unchanged by the gauging. This relies on the exchange of symmetry group extensions with mixed anomalies \cite{Tachikawa:2017gyf}. In particular, when the self-duality conditions \eqref{eq:duality_condition} are satisfied, the symmetries and anomalies are the same after gauging.

For simplicity, in this work we have focused on the case of a single isometry, associated with one $S^1$ fibre of the target space. In general, a manifold $M$ can have many isometries, with Killing vectors that may or may not commute. As a result the isometry symmetry can be greater than $\U(1)$ and, for example, can be non-Abelian. It would be interesting to carry out a similar analysis to the one of this work for those cases. The case of several commuting Killing vectors is a straightforward generalisation, and the pure 't Hooft anomalies in the case where the global symmetry is continuous, i.e.~$\U(1)^d$, have been discussed in \cite{Hull:1989jk,Hull:1991uw}. When the Killing vectors do not commute, so the isometry symmetry is non-Abelian, the situation is less clear.
Firstly, the identification of the dual isometry symmetry is not known in this case, primarily since non-abelian T-duality is less well understood than the Abelian version. Secondly, if the non-Abelian symmetry is discrete, the cocycle technology that we have employed in this work will not apply since singular cohomology groups can only have coefficients in Abelian groups and a more categorical description of the symmetry structure would be required. The situation may be clearer when the target space is a semi-simple Lie group $G$ and the NLSM has $G\times G$ isometry symmetry, since $H^2(G)$ is trivial and the WZ term is guaranteed to preserve the continuous symmetry.

Our initial motivation for the present work was related to a non-invertible symmetry in these models, and many of the puzzles that remain are also related to this. 
The first question concerns the identification of a larger set of non-invertible defects. In \cite{Arias-Tamargo:2025xdd}, the half-space gauging procedure was used to construct the defect, and the discrete symmetry which was gauged was a subgroup of the isometry symmetry. The main reason for that was that the doubled torus construction developed in the context of T-duality \cite{Hull2006} could be readily adapted to perform this gauging. We have now seen that generically these models have an additional dual isometry symmetry with a non-anomalous discrete subgroup. This symmetry can also be gauged and possibly used to construct other non-invertible defects. In the example of the compact boson, gauging different combinations of the momentum and winding symmetries leads to non-invertible defects at any rational value of the radius squared: $2\pi R^2=p/q$. A natural guess is that in a more general NLSM gauging a $\Z_p\times \Z_q$ subgroup of the isometry and dual isometry symmetries will also lead to non-invertible defects. The results of the present work will be essential to this construction in order to avoid both pure and mixed 't Hooft anomalies.
There will again be self-duality conditions which must be satisfied for the theory to admit a non-invertible defect. For the case of a single isometry, it would be natural for these to be a simple generalisation of \eqref{eq:duality_condition}: $2\pi G=p/q$ and $pF=q\widetilde{F}$. 
It would be interesting to check this explicitly, and we hope to be able to report on this soon.

These non-invertible symmetries have many applications, which we also hope to study. First and foremost, there are Ward identities associated with them, just as with invertible symmetries. The analysis of this work will be useful for that purpose because the Ward identities depend on the charges of vertex operator insertions.
Therefore, having identified the symmetries and how they are modified by the discrete gauging should streamline that analysis.

Finally, we comment on the role of conformal symmetry in our work. While much of the literature about non-invertible symmetries in two dimensions discusses rational CFTs, conformal symmetry does not play any role in our approach. Instead, our construction relies solely on geometric properties of the target space. Indeed, there are interesting examples of non-conformal NLSMs admitting non-invertible defects. In these cases, when the symmetry is preserved along an RG flow it can be used to constrain the IR fixed point of the flow. We also intend to report on these applications in the near future.

\section*{Acknowledgments}

It is a pleasure to thank Riccardo Argurio, Giovanni Galati, Chris Hull, Jeremias Aguilera-Damia, and Rishi Mouland for interesting discussions.
GAT is supported by the STFC Consolidated Grants ST/T000791/1 and ST/X000575/1.
MVCH is supported by a President's Scholarship from Imperial College London.

\begin{appendix}

\section{Quantisation of \texorpdfstring{$\iota_k v$}{k.v} and the Chern-Simons level}
\label{app:quantisation}

In this appendix, we discuss the quantisation of $\iota_k v$ in the case where there is a $\U(1)$ isometry symmetry, such that $v$ is a globally defined 1-form on $M$ satisfying \eqref{eq:iH_exact}. In particular, we focus on the case where $[F]\neq0$ so the bundle \eqref{eq:M_bundle} is non-trivial. If $[F]=0$ then, as shown in subsection~\ref{subsec:anomalies_U(1)}, there always exists a globally defined choice of $v$ for which $\iota_k v=0$ so its quantisation is unimportant. When $[F]\neq0$, such a choice of $v$ generically does not exist and the quantisation of $\iota_k v$ is then non-trivial.

We can decompose $v$ into a basic and fibre component as in \eqref{eq:v=v'+fibre},
\begin{equation}
    v = v' + (\iota_k v) \dd{X} \,,
\end{equation}
where we are using coordinates \eqref{eq:k_local}.
It follows from \eqref{eq:Lv=0} that $\iota_k v$ is a constant, so $\dd{v'} = \dd{v}$. Since $v$ is a globally defined 1-form, under a coordinate transformation of the fibre direction $X \to X - f$ (where $f\sim f+2\pi$), $v'$ must transform as $v' \to v' + (\iota_k v) \dd{f}$.
In particular, if $\iota_k v$ is an integer then $v'$ transforms as a conventionally normalised $\U(1)$ connection on the base $N$ with curvature $\widetilde{F} = \dd{v'} = \dd{v}$.

Now, recall from section~\ref{sec:NLSMs} that the WZ term of the NLSM \eqref{eq:NLSM_action} only gives a well-defined two-dimensional theory if $H/2\pi$ represents an integral cohomology class. The closed 3-form $H$ can be decomposed as in \eqref{eq:H_split}, which we repeat here for convenience:
\begin{equation}
    H = \bar{H} + \frac{1}{2\pi} \widetilde{F} \wedge \xi \,.
\end{equation}
Here, $\bar{H}/2\pi$ is a basic 3-form with integer periods. Then $H/2\pi$ has integer periods if and only if $\widetilde{F} \,\wedge\, \xi / (2\pi)^2$ does. Let $\gamma_k$ be the 1-cycle generated by $k$, then $\int_{\gamma_k} \xi = 2\pi$ from \eqref{eq:xi=A+dX}. Since $\widetilde{F}$ is basic, in order for $H/2\pi$ to have integral periods we require
\begin{equation}
    \frac{1}{2\pi} \int_{\Sigma_2} \widetilde{F} \in \Z
\end{equation}
for all 2-cycles $\Sigma_2$ in $M$. That is, we require that the curvature $\widetilde{F}$ of the connection $v'$ is conventionally normalised. This is the case if and only if the gauge transformations of the connection $v'$ are $v' \to v' + \dd\lambda'$ with $\lambda' \sim \lambda' + 2\pi\Z$. 

We saw above that $v'$ transforms in this manner with $\lambda' = (\iota_k v) f$ for a $2\pi$-periodic function $f$. Therefore, $H/2\pi$ only has integer periods if $\iota_k v \in \Z$. Therefore, the condition imposed on $H$ for the WZ term to be well-defined already implies that $\iota_k v\in\Z$.

This result is relevant here since $\iota_k v$ is the coefficient of the Chern-Simons theory \eqref{eq:pure_inflow} which describes the inflow theory of the pure isometry symmetry anomaly.
An anomaly inflow theory should be such that if it is placed on a \emph{closed} 3-manifold $V$ then it gives a well-defined topological theory. 
For Abelian Chern-Simons theory, the standard normalisation of the action is 
\begin{equation}
    S_{\text{CS}} =\frac{K}{4\pi} \int_V C\wedge dC \,.
\end{equation}
Then $e^{iS_{\text{CS}}}$ is gauge-invariant on arbitrary oriented closed 3-manifolds $V$ provided that the level $K$ is an even integer. Comparing to \eqref{eq:pure_inflow}, the level in our case is $K=2\iota_k v$ so the inflow theory is well-defined on arbitrary closed 3-manifolds if $\iota_k v \in\Z$.

In particular, we do not need to demand that $\iota_k v\in\Z$ as an independent constraint such that the inflow theory \eqref{eq:pure_inflow} is well-defined. Instead, the quantisation of $\iota_k v$ follows from the original data of the theory.

\section{Exact sequences for group extensions} \label{app:bockstein}

In this appendix, we include some background material regarding group extensions, which we use frequently in the main text.

\subsection{Short exact sequence and group cohomology}

Given two Abelian groups $A$ and $H$, we want to study the possible extensions $G$ such that
\begin{align}
    H=G/A\,.
\end{align}
Since $A$ must be a normal subgroup of $G$, we can always construct a short exact sequence
\begin{align}\label{eq:SES_appendix}
    1\to A\xrightarrow{\iota} G\xrightarrow{\pi} H\to 1\,.
\end{align}
An exact sequence of groups is a sequence of group homomorphisms in which the image of each map is the kernel of the next. A short exact sequence is an exact sequence of the form \eqref{eq:SES_appendix}. Exactness implies that $\iota:A\to G$ is an injection and that $\pi:G \to H$ is a surjection.
If $\iota(A)\subseteq Z(G)$ the extension is called \emph{central} and the possible extensions \eqref{eq:SES_appendix} are classified by group cohomology $H^2_{\text{grp}}(H,A)$, as we now review. Here we consider only discrete groups. When the groups are continuous this algebraic description of the group cohomology is no longer applicable and should instead be defined in terms of the singular cohomology of the classifying space of $H$ \cite{Hatcher}.

Given a group cohomology class $[\sigma_2] \in H^2_{\text{grp}}(H,A)$, a representative $\sigma_2$ is a map
\begin{align}\label{eq:sigma2}
    \sigma_2:H\times H \to A
\end{align}
which satisfies $\delta\sigma_2=0$ where the coboundary operator is defined by
\begin{equation}
    (\delta \sigma_2)(h_1,h_2,h_3) = \sigma_2(h_2,h_3) - \sigma_2(h_1 h_2,h_3) + \sigma_2(h_1,h_2 h_3) - \sigma_2(h_1,h_2) \,.
\end{equation}
The group $G$ can then be defined as the set of pairs $(a,h)\in A\times H$ endowed with a group product defined in terms of $\sigma_2$:
\begin{align}
    (a_1,h_1)\cdot (a_2,h_2)=(a_1+a_2+\sigma_2(h_1,h_2), h_1 + h_2)\,.
\end{align}
Here we use additive notation for the group operation in $A$ and $H$ as they are Abelian, even if the extension $G$ generically is not. One can then verify that this definition does not depend on the choice of representative of $[\sigma_2]$.

In the other direction, given a central group extension, one way to find a representative $\sigma_2$ is as follows. Given the sequence \eqref{eq:SES_appendix} we can choose any map $\sigma_1:H\to G$ and define
\begin{align}
\begin{split}
    \sigma_2: H\times H&\to G\\ \sigma_2(h_1,h_2)&=\sigma_1(h_1)\cdot\sigma_1(h_2)\cdot\left(\sigma_1(h_1+h_2)\right)^{-1}\,.
\end{split}
\end{align} 
One can check that $\sigma_2(h_1,h_2)\in A\subseteq G$ so the map can be understood as $\sigma_2 : H \times H \to A$. Moreover, making a different choice of section $\sigma_1$ does not change the cohomology class $[\sigma_2]$.

We note that $\sigma_1$ is not required to be a group homomorphism. If it is a homomorphism then $\sigma_2$ is exact, $[\sigma_2]=[0]\in H^2_{\text{grp}}(H,A)$, and we say that the short exact sequence \eqref{eq:SES_appendix} splits. Split central extensions are automatically trivial, i.e.~$G\cong A\times H$.

\subsection{The snake lemma and the Bockstein homomorphism}
\label{app:snake}

In the main text, we make frequent use of the Bockstein homomorphism, which appears as a discrete analogue of the curvature of a connection. We now review its definition, which comes associated to a short exact sequence of the form \eqref{eq:SES_appendix} and stems from a result known as the snake lemma. The lemma states that associated to a short exact sequence of Abelian groups, there is a long exact sequence of singular cohomology groups (on any given space $M$) with coefficients in the groups which make up the short exact sequence. A sketch of the argument is as follows.

Recall that the short exact sequence \eqref{eq:SES_appendix} is defined by a pair of maps $\iota$ and $\pi$.
These induce maps on singular cochains on $M$. In particular, $\iota: A \to G$ induces a map $f:C^n(M,A)\to C^{n}(M,G)$ and $\pi: G \to H$ induces a map $g:C^n(M,G)\to C^{n}(M,H)$. We also have the coboundary operator $\delta:C^n(M,\cdot)\to C^{n+1}(M,\cdot)$. One can then build the following commutative diagram,
\begin{align}\label{eq:snake_lema}
    \begin{matrix}
        & & H^n(M,A) & \to & H^n(M,G) & \to & H^n(M,H) & \textcolor{red}{\to}  & \\
        & & \downarrow\Scale[0.75]{\iota'} & & \downarrow\Scale[0.75]{\iota'} & & \downarrow\Scale[0.75]{\iota'} &  &  \\
        & & C^n(M,A)   & \xrightarrow{f} & C^n(M,G)   &  \xrightarrow{g} & C^n(M,H) & \xrightarrow{}& 1 \\
        &&\downarrow \Scale[0.75]{\delta}&&\downarrow\Scale[0.75]{\delta}&&\downarrow\Scale[0.75]{\delta}&&  \\
        1 &\xrightarrow{}&Z^{n+1}(M,A)&\xrightarrow{f}&Z^{n+1}(M,G)&\xrightarrow{g}&Z^{n+1}(M,H)& &  \\    &&\downarrow\Scale[0.75]{\pi'}&&\downarrow\Scale[0.75]{\pi'}&&\downarrow\Scale[0.75]{\pi'}&&  \\
        &\textcolor{red}{\to}&H^{n+1}(M,A)&\xrightarrow{f}&H^{n+1}(M,G)&\xrightarrow{g}&H^{n+1}(M,H)&&       
    \end{matrix}
\end{align}
Here, $Z^n(M,\cdot)$ denotes closed cochains, $\iota'$ is the natural inclusion, and $\pi'$ the natural projection obtained by taking the quotient of $Z^{n+1}(M,\cdot)$ by the group of exact cochains. 

The Bockstein homomorphism is the connecting group homomorphism from the RHS of the top row of \eqref{eq:snake_lema} to the LHS of the bottom row (i.e.~corresponding to the red arrow); that is, 
\begin{align}
    \text{Bock}: H^n(M,H)\to H^{n+1}(M,A)\,.
\end{align}
It can be used to construct a long exact sequence of cohomology groups:
\begin{align}
    \cdots\to H^{n}(M,A)\to H^{n}(M,G)\to H^{n}(M,H)\xrightarrow{\text{Bock}} H^{n+1}(M,A)\to\cdots
\end{align}
The Bockstein is constructed as follows.
\begin{enumerate}
    \item We begin with a class $[\omega]\in H^n(M,H)$ and consider a representative $\omega$. This can, of course, be seen as a closed $n$-cochain $\omega \in C^n(M,G)$ with $\delta\omega=0$ via the inclusion $\iota'$.
    \item Because the second row in \eqref{eq:snake_lema} is exact, $g$ is surjective, so there is some $\widetilde{\omega}\in C^n(M,G)$ such that $\omega=g(\widetilde{\omega})$.
    \item We now consider $\delta\widetilde{\omega}$. Since the diagram commutes, $g(\delta \widetilde{\omega}) = \delta(g(\widetilde{\omega})) = \delta\omega = 0$, so $\delta \widetilde{\omega} \in \ker(g)$.
    \item Exactness of the third row of \eqref{eq:snake_lema} implies $\delta \widetilde{\omega} = f(\nu)$, for some $\nu\in Z^{n+1}(M,A)$. Indeed, the fact that $\nu$ is closed follows from $f(\delta\nu) = \delta (f(\nu)) = \delta^2 \widetilde{\omega} = 0$ as $f$ is injective.
    \item Finally, since $\nu$ is closed, it represents a cohomology class $[\nu]\in H^{n+1}(M,A)$.
\end{enumerate}
This procedure defines a map 
\begin{align}
    \text{Bock}([\omega])=[\nu]\,,
\end{align}
and one can check that the choices involved (i.e.~the choice of representative $\omega$ and the choice of lift $\widetilde{\omega}$) only change $\nu$ by the addition of an exact cochain, so the map between cohomology classes is well-defined. This map is the Bockstein homomorphism.

\subsection{The Bockstein for the sequence of cyclic groups}
\label{app:cyclic_bock}

We now discuss the Bockstein homomorphism associated with the short exact sequence
\begin{align}\label{eq:ZN_SES}
    1\xrightarrow{}\Z_p\xrightarrow{\times q} \Z_{pq} \xrightarrow{\text{mod }q} \Z_q\to 1 \,.
\end{align}
This sequence appears several times in the main text with different values of $p$ and $q$. We remark that this sequence splits when $\gcd(p,q)=1$. Indeed, in this case one can explicitly construct a group homomorphism $\sigma_1:\Z_q \to \Z_{pq}$ with $\pi\circ\sigma_1 = \text{id}$. It is $\sigma_1(n) = p p^{-1} n \text{ mod }pq$, where $p^{-1}$ is the multiplicative inverse of $p$ modulo $q$ (which is well-defined since $p$ and $q$ are coprime).

The Bockstein for this sequence is a map
\begin{equation}\label{eq:bock_def}
    \text{Bock}:H^n(M,\Z_q)\to H^{n+1}(M,\Z_p)
\end{equation}
and can be written explicitly as
\begin{align}\label{eq:Bock_cyclic_groups_app}
    \text{Bock}([\omega_q])= \left[ \left(\frac{1}{q} \,\delta\widetilde{\omega}_{pq} \right)\text{ mod } p \right]\,.
\end{align}
The logic of the snake lemma in this situation is as follows. Given $[\omega_q] \in H^n(M,\Z_q)$, we can regard a representative $\omega_q$ as a map that gives an integer defined mod $q$ such that 
\begin{equation}\label{eq:domega}
    \delta \omega_q = 0\text{ mod } q \,.
\end{equation}
The lift to the $\Z_{pq}$-valued cochain $\widetilde{\omega}_{pq}$ consists of simply seeing this map as giving us the same integer defined mod $pq$. It follows from \eqref{eq:domega} that $\frac{1}{q} \delta \widetilde{\omega}_{pq}$ gives an integer defined mod $p$. 
Indeed, a representative of the RHS of \eqref{eq:Bock_cyclic_groups_app} is $\nu\in Z^{n+1}(M,\Z_p)$ such that when multiplying by $q$ and looking at the result mod $pq$, it gives precisely $\delta \widetilde{\omega}_{pq}$ mod $pq$.

In \eqref{eq:int_lift_bock} we use another expression for the Bockstein \eqref{eq:bock_def} for $n=1$, written in terms of a $\U(1)$ gauge field. We now discuss this relation following \cite{Brennan:2023mmt} (see also \cite{Cordova:2018acb}).
Consider the commuting diagram
\begin{equation}\label{eq:SES_lift}
\begin{tikzpicture}[descr/.style={fill=white,inner sep=1.5pt}]
        \matrix (m) [
            matrix of math nodes,
            row sep=1.5em,
            column sep=2em,
            text height=1.5ex, text depth=0.25ex
        ]
        { 1 & \Z & \,\Z\, & \Z_q & 1 \\ 1 & \Z_p & \Z_{pq} & \Z_q & 1 \\
        };

        \path[overlay,->]
        (m-1-1) edge (m-1-2)
        (m-1-2) edge node[label=above:$\times q$]{} (m-1-3)
        (m-1-3) edge node[label=above:mod $q$]{} (m-1-4)
        (m-1-4) edge (m-1-5)
        (m-2-1) edge (m-2-2)
        (m-2-2) edge node[label=below:$\times q$]{} (m-2-3)
        (m-2-3) edge node[label=below:mod $q$]{} (m-2-4)
        (m-2-4) edge (m-2-5)
        (m-1-2) edge node[label=right:mod $p$]{} (m-2-2)
        (m-1-3) edge node[label=right:mod $pq$]{} (m-2-3)
        (m-1-4) edge node[label=right:id]{} (m-2-4);
\end{tikzpicture}
\end{equation}
Both the top and bottom rows are short exact sequences, so there is an associated commuting diagram
\begin{equation}
\begin{tikzpicture}[descr/.style={fill=white,inner sep=1.5pt}]
        \matrix (m) [
            matrix of math nodes,
            row sep=1.5em,
            column sep=1em,
            text height=1.5ex, text depth=0.25ex
        ]
        { \cdots & H^1(M,\Z) & \,H^1(M,\Z)\, & H^1(M,\Z_q) & H^2(M,\Z) & \cdots \\ \cdots & H^1(M,\Z_p) & H^1(M,\Z_{pq}) & H^1(M,\Z_q) & H^2(M,\Z_p) & \cdots \\
        };

        \path[overlay,->]
        (m-1-1) edge (m-1-2)
        (m-1-2) edge (m-1-3)
        (m-1-3) edge (m-1-4)
        (m-1-4) edge node[label=above:$\widehat{\Bock}$]{} (m-1-5)
        (m-1-5) edge (m-1-6)
        (m-2-1) edge (m-2-2)
        (m-2-2) edge (m-2-3)
        (m-2-3) edge (m-2-4)
        (m-2-4) edge node[label=above:$\Bock$]{} (m-2-5)
        (m-2-5) edge (m-2-6)
        (m-1-2) edge node[label=right:mod $p$]{} (m-2-2)
        (m-1-3) edge node[label=right:mod $pq$]{} (m-2-3)
        (m-1-4) edge node[label=right:id]{} (m-2-4)
        (m-1-5) edge node[label=right:mod $p$]{} (m-2-5);
\end{tikzpicture}
\end{equation}
Here, we have denoted the Bockstein homomorphism associated with the sequence on the top row of \eqref{eq:SES_lift} by $\widehat{\Bock}$ to distinguish it from the one associated to the sequence \eqref{eq:ZN_SES} in the bottom row of \eqref{eq:SES_lift}. Commutativity implies
\begin{equation}\label{eq:bock_lift}
    \Bock = \widehat{\Bock} \text{ mod } p \,.
\end{equation}
That is, the Bockstein associated with the sequence \eqref{eq:ZN_SES} can be understood as a reduction of a $\Z$-valued Bockstein modulo $p$. The construction of the $\Z$-valued Bockstein $\widehat{\Bock}$ is completely analogous to the expression discussed in \eqref{eq:Bock_cyclic_groups_app}, and we find
\begin{equation}\label{eq:int_bock_work}
    \widehat{\Bock}([\omega_q]) = \left[ \frac{1}{q} \delta \widehat{\omega}_\Z \right] \,,
\end{equation}
where $\widehat{\omega}_\Z \in C^1(M,\Z)$ is an integer cochain such that $\omega_q = \widehat{\omega}_\Z \text{ mod } q$. As above, it follows from $\omega_q \in H^1(M,\Z_q)$ that $\delta \widehat{\omega}_\Z$ is a multiple of $q$, so the right-hand side of \eqref{eq:int_bock_work} is an integer class.
From \eqref{eq:bock_lift}, we then have
\begin{equation}\label{eq:bock_lift_2}
    \Bock([\omega_q]) = \left[ \left(\frac{1}{q} \delta \widehat{\omega}_\Z\right)  \text{ mod } p\right] \,.
\end{equation}
Indeed, the equivalence between this expression and \eqref{eq:Bock_cyclic_groups_app} is straightforward to demonstrate. The relevance of the integer Bockstein is that we can interpret $\widehat{\omega}_\Z$ as encoding the holonomies of a $\U(1)$ connection $\widehat{A}$:
\begin{equation}\label{eq:int_lift_better}
    \int_\gamma \widehat{A} = \frac{2\pi}{q} \widehat{\omega}_\Z (\gamma) \,,
\end{equation}
where $\gamma$ is a 1-cycle. Such a connection then has $\exp(i\int_\gamma \widehat{A}) = \exp(2\pi i n/q)$ where $n = \widehat{\omega}_\Z(\gamma) \in \Z$ and so has the holonomies of a $\Z_q$ gauge field. 
Roughly speaking, the relation \eqref{eq:int_lift_better} between the $\U(1)$ connection $\widehat{A}$ and the integral cochain $\widehat{\omega}$ can be written
\begin{equation}
    \widehat{A} \sim \frac{2\pi}{q} \widehat{\omega}_\Z \,.
\end{equation}
We use the symbol $\sim$ to indicate that this is not a rigorous equation and the objects on either side are not of the same type. This expression, which appeared already in \eqref{eq:int_lift_relation}, should be understood as encoding the holonomies of $\widehat{A}$ as in \eqref{eq:int_lift_better}.
On trivial cycles, $\gamma = \partial D$, we have
\begin{equation}
    \int_D \widehat{F} = \frac{2\pi}{q} \delta \widehat{\omega}_\Z (D) \,,
\end{equation}
so the curvature of the $\U(1)$ connection satisfies
\begin{equation}
    \frac{1}{2\pi} \widehat{F} = \frac{1}{q} \delta \widehat{\omega}_\Z \in H^2(M,\Z) \,.
\end{equation}
The expression \eqref{eq:bock_lift_2} can then be written in terms of this curvature as
\begin{equation}
    \Bock([\omega_q]) = \left[ \frac{\widehat{F}}{2\pi} \text{ mod }p \right] \,.
\end{equation}
This is the relation required in \eqref{eq:int_lift_bock}.

\end{appendix}

\bibliographystyle{JHEP}
\bibliography{refs}

\end{document}